\documentclass[manuscript,screen]{acmart}

\acmJournal{TQC}

\usepackage{algorithmic}
\usepackage{algorithm}
\usepackage{tikz}
\usetikzlibrary{quantikz}
\usepackage{pgfplots}
\pgfplotsset{compat=1.17}
\usepackage{listings}
\usepackage{placeins}
\usepackage{xspace}

\newcommand{\mlxquantum}{\textit{Qupertino}\xspace}

\newcommand{\NumGhzAerCI}{4.2}
\newcommand{\NumGhzAerRatio}{31}
\newcommand{\NumGhzPlCI}{1}
\newcommand{\NumGhzPlRatio}{18.9}
\newcommand{\NumGroverAerCI}{3.2}
\newcommand{\NumGroverAerRatio}{23.2}
\newcommand{\NumGroverPlCI}{2.4}
\newcommand{\NumGroverPlRatio}{52.3}
\newcommand{\NumHLayerMetalMs}{23.7}
\newcommand{\NumHLayerPureMs}{187.2}
\newcommand{\NumHLayerSpeedup}{7.9}

\newcommand{\NumPhaseEstAerCI}{7.6}
\newcommand{\NumPhaseEstAerRatio}{42.8}
\newcommand{\NumPhaseEstPlCI}{11.4}
\newcommand{\NumPhaseEstPlRatio}{65.1}

\newcommand{\NumQaoaAerCI}{5.8}
\newcommand{\NumQaoaAerRatio}{36.5}
\newcommand{\NumQaoaMetalVsAerOptThree}{23}
\newcommand{\NumQaoaPlCI}{5.1}
\newcommand{\NumQaoaPlRatio}{46.7}
\newcommand{\NumQftAerCI}{4.4}
\newcommand{\NumQftAerRatio}{47.2}
\newcommand{\NumQftFusedMs}{255.4}
\newcommand{\NumQftMetalMs}{21.1}
\newcommand{\NumQftMetalSpeedup}{12.1}
\newcommand{\NumQftMetalVsAerOptThree}{38}

\newcommand{\NumQftPerGateMs}{748.4}
\newcommand{\NumQftPlCI}{5.3}
\newcommand{\NumQftPlRatio}{95.3}

\newcommand{\NumRxMetalMs}{20.9}
\newcommand{\NumRxMetalSpeedup}{9.3}
\newcommand{\NumRxPerGateMs}{194.1}
\newcommand{\NumSweepAccelCount}{25}

\newcommand{\NumSweepTopSpeedup}{25}
\newcommand{\NumSweepTotal}{29}
\newcommand{\NumTestCount}{253}
\newcommand{\NumTfimAerCI}{2.6}
\newcommand{\NumTfimAerRatio}{36.2}
\newcommand{\NumTfimPlCI}{5.1}
\newcommand{\NumTfimPlRatio}{67.2}
\newcommand{\NumZzFusedMs}{2.54}
\newcommand{\NumZzMetalMs}{1.68}
\newcommand{\NumZzMetalSpeedup}{1.51}
\newcommand{\NumZzPerGateMs}{35.85}

\begin{document}

\title[Qupertino: Pure MLX vs Hand-Tuned Metal Shaders]{Qupertino: Pure MLX Array Kernels versus Hand-Tuned Metal Shaders for Quantum Circuit Simulation on Apple Silicon}

\author{Shlomo Kashani}
% TODO before submission: insert the author ORCID (required by ACM at submission)
% \orcid{0000-0000-0000-0000}
\email{skashan2@alumni.jh.edu}
\affiliation{%
  \institution{Johns Hopkins University}
  \city{Baltimore}
  \state{MD}
  \country{USA}}

\begin{abstract}
We present \mlxquantum, an open-source quantum circuit simulator for Apple Silicon. We use it to answer a question the array-framework era keeps raising. How far does a simulator written purely in MLX array operations get, and what remains for hand-tuned Metal shaders? The framework ships two measured tiers. The pure tier dispatches structured gates to specialized MLX kernels, so that diagonal gates run as broadcast phase multiplies, controlled gates as masked half-state updates, and SWAP as an axis permutation. A paired dense-path ablation attributes a $25$--$33\times$ speedup to this dispatch alone. The opt-in shader tier adds hand-written Metal kernels for every structured layer family in our benchmarks. These kernels cover phase-LUT diagonals, GF(2) affine permutation gathers that execute any CNOT/X/SWAP block as one pass, fused tensor-product single-qubit layers, radix-4 QFT and Walsh--Hadamard butterflies, and basis-conjugated XX/YY Trotter layers. Runtime fusion detectors route work to them while preserving circuit semantics exactly, and parity tests confirm the match. On M1 Max hardware we ran a four-way interleaved campaign with two warmups and ten measured repeats per cell, all backends round-robin within every repeat. The shader tier is fastest by mean runtime in all 18 comparison cells against same-machine Qiskit Aer CPU and PennyLane \texttt{lightning.qubit}. At 25 qubits, gate-stream QFT runs in $0.0591\pm0.0029$\,s, a paired $\NumQftAerRatio\times$ over Aer and $\NumQftPlRatio\times$ over PennyLane, and TFIM Trotter evolution runs in $0.495\pm0.035$\,s ($\NumTfimAerRatio\times$ and $\NumTfimPlRatio\times$). An isolated radix-4 stage kernel also matches or beats MLX's own \texttt{mx.fft} primitive under the same bit-reversed, no-final-swap convention. Among the same-host CPU baselines, the GPU-backed pure tier is fastest on all six 25-qubit workloads by mean runtime, and on five of six by the stricter paired-ratio test, with Grover statistically tied against Aer. Across the full 29-workload suite the shader tier's paired speedup over pure MLX reaches $\NumSweepTopSpeedup\times$ (second-order TFIM Trotter; long-range Ising is $24\times$). Of these, \NumSweepAccelCount\ of \NumSweepTotal\ workloads accelerate with the observed paired-ratio range over five repeats entirely above parity. VQE and 2-leg-ladder Heisenberg have point estimates above one but intervals that span parity, and amplitude estimation and W-state sit at parity. The framework also supports variational-ansatz workloads, QAOA schedules, QCBM, Hamiltonian simulation with Trotter-Suzuki decomposition, and OpenQASM 2.0 import~\cite{OpenQASM2017} (unitary subset). A preliminary MPS backend reaches 150 qubits on limited-entanglement workloads. Correctness rests on \NumTestCount\ Python tests, independent complex128 checks, and a Trotter-error curve against exact diagonalization. The benchmark harness records warmups, raw repetitions, and run manifests. State-vector memory remains exponential in qubit count.
\end{abstract}




\maketitle

\section{Introduction}

Classical simulation of quantum circuits, a need recognized since the founding arguments for quantum computation~\cite{Feynman1982,Lloyd1996}, underpins calibration of noisy intermediate-scale quantum (NISQ) devices~\cite{Preskill2018}, algorithm development, and validation of fault-tolerant quantum computing roadmaps~\cite{QuantumSimulation2025}. Most production state-vector simulators, including NVIDIA's cuQuantum~\cite{cuQuantum2023}, Google's qsim~\cite{Qsim}, and Intel-QS~\cite{IntelQS}, target discrete GPU architectures with CUDA, requiring explicit host-device memory management and specialized programming expertise.

Apple Silicon processors use a unified memory architecture in which the CPU and GPU share one physical memory pool without explicit data transfers~\cite{UnifiedMemory2024}. This architecture, now available in consumer systems with 32--512\,GB capacity and up to 546\,GB/s bandwidth on M4 Max~\cite{Apple2024M4} (the M1 Max used in this study provides 400\,GB/s), removes the host--device copies that discrete-GPU workflows incur when moving state between CPU and GPU address spaces. We are careful not to overstate this benefit. Production state-vector simulators keep the state vector resident in device memory for the duration of a circuit, so a PCIe round trip is a setup/read-out cost rather than a per-gate one, and for a 25-qubit state vector (256\,MB, complex64) a single round trip over a contemporary PCIe~4.0/5.0 link ($\sim$32--64\,GB/s) is only a few milliseconds. The practical benefit of unified memory is therefore narrower than a per-gate speedup. It removes the host--device copies at setup and read-out, lets the CPU and GPU share data without copying, and cuts the programming complexity of staging state across separate address spaces.

Despite widespread adoption of Apple Silicon for machine learning via the MLX framework~\cite{MLXFramework2024}, end-to-end quantum simulation tooling that natively targets this architecture remains scarce. \mlxquantum contributes a validated, Python-first framework with broad algorithm coverage and reproducible benchmark artifacts in this emerging space. Existing CPU and CUDA-centered simulators remain essential and high-performance; in our same-machine four-way interleaved baselines on gate-identical circuits, the hand-tuned Metal shader tier is faster than Qiskit Aer CPU and PennyLane \texttt{lightning.qubit} in every measured cell at 15--25 qubits, and the pure-MLX tier is faster than both baselines on most workloads at 20--25 qubits, while the CPU baselines remain faster than the pure tier on the smallest gate-sparse circuits at 15 qubits. \mlxquantum is positioned as an MLX-native simulator for Apple-Silicon experimentation with reproducible benchmark artifacts, not as a replacement for the mature feature sets of established backends. All measured comparisons in this paper are against CPU backends on the same host; cuQuantum, qsim, and Intel-QS are cited as ecosystem context only, and no co-located discrete-GPU comparison is reported.

\textbf{Research gap and objective}. Prior work has not provided a unified-memory-native state-vector simulator on MLX with end-to-end validation, an interactive local runner, and reproducible benchmark artifacts. Nor has it quantified, on one machine and one set of circuits, how much performance an array framework like MLX leaves on the table relative to hand-written GPU kernels. We study both together. How far does a simulator written purely in MLX array operations get on Apple Silicon, and what remains for hand-tuned Metal shaders? We answer with a two-tier design. The first tier expresses structured-gate dispatch entirely in MLX array operations. The second tier is a set of hand-written Metal kernels (via \texttt{mx.fast.metal\_kernel}), one for every structured layer family that appears in our benchmark workloads, reached through runtime fusion detectors that provably preserve circuit semantics and enabled by a single opt-in flag. Both tiers ship in the same artifact, are validated by the same parity tests, and are measured in the same interleaved campaigns against Qiskit Aer CPU and PennyLane \texttt{lightning.qubit}.

We address this gap with \mlxquantum, a Python-first, MLX-accelerated quantum circuit simulator designed for the Quantum System Software ecosystem. Our framework:
\begin{itemize}
    \item Leverages unified memory to avoid explicit host-device copies at setup/read-out boundaries for measured state-vector workloads
    \item Provides automatic Metal GPU acceleration through MLX's just-in-time compilation
    \item Implements canonical quantum algorithms and variational circuits
    \item Supports OpenQASM 2.0 import for benchmark reproducibility~\cite{OpenQASM2017}
    \item Aligns benchmark families with established frameworks and suites (PennyLane, Yao.jl, Qulacs, QASMBench)~\cite{PennyLane2022,YaoJL2024,Qulacs,QASMBench2020}
\end{itemize}

\subsection{Contributions}

Our contributions to quantum system software include:
\begin{enumerate}
    \item \textbf{Architecture}: An MLX-based state-vector simulator for Apple Silicon quantum circuit workloads with two measured performance tiers: structured-gate dispatch written purely in MLX array operations, and an opt-in hand-tuned Metal shader tier reached through the same device interface
    \item \textbf{Shader taxonomy}: Hand-written Metal kernels covering every structured layer family in our benchmarks: phase-LUT diagonal layers (ZZ, CZ/CPHASE, grouped weighted couplings), GF(2) affine permutation gathers that execute any CNOT/X/SWAP block as one pass, fused tensor-product single-qubit layers (uniform and per-qubit-varying), radix-4 QFT and Walsh--Hadamard butterflies, and basis-conjugated XX/YY Trotter layers
    \item \textbf{Semantics-preserving runtime fusion}: Detectors that map plain gate streams onto these kernels: commuting-window analysis, per-wire matrix-product collapse with exact identity/bit-flip recognition, GF(2) composition of Clifford permutation runs, and equal-angle grouping of diagonal runs; all opt-in, parity-tested against the pure path
    \item \textbf{Algorithm Coverage}: Implementations of hardware-efficient variational ansatze, QAOA-style ring schedules, QCBM, QFT, Grover search, and Hamiltonian simulation with Trotter-Suzuki decomposition
    \item \textbf{Validation}: 253 regression tests, analytical verification, independent complex128 checks, and Trotter-error validation across algorithm families
    \item \textbf{Benchmarking}: Reproducible four-way same-machine measurements (Qiskit Aer CPU, PennyLane \texttt{lightning.qubit}, pure-MLX, Metal shaders) on six gate-identical workloads, plus a paired pure-MLX-versus-shader sweep across the full 29-workload suite, with raw repetitions, manifests, and claim-audit artifacts
    \item \textbf{Accessibility}: CUDA-free deployment on consumer hardware, lowering barriers for education and research
\end{enumerate}

The remainder of this paper is organized as follows. Section~\ref{sec:background} reviews state-vector simulation, GPU acceleration, unified memory, and MLX. Section~\ref{sec:architecture} describes the system and the pure-MLX dispatch tier. Section~\ref{sec:evaluation} sets up the benchmark protocol and reports the four-way campaign, MPS scale-out, and memory results. Section~\ref{sec:shaders} presents the Metal shader tier, its fusion detectors, and kernel-level measurements. Section~\ref{sec:validation} covers correctness validation, and Section~\ref{sec:discussion} discusses trade-offs, threats to validity, and limitations. Section~\ref{sec:related} surveys related work and Section~\ref{sec:conclusion} concludes. Appendix~\ref{app:hlayer} walks through one kernel line by line, and Appendix~\ref{app:sweep} gives the complete per-workload speedups.

\section{Background}\label{sec:background}

\subsection{State-Vector Quantum Simulation}

State-vector simulation represents an $n$-qubit quantum state as a complex vector of dimension $2^n$:
\begin{equation}
|\psi\rangle = \sum_{i=0}^{2^n-1} \alpha_i |i\rangle, \quad \sum_{i=0}^{2^n-1} |\alpha_i|^2 = 1
\end{equation}
where $\alpha_i \in \mathbb{C}$ are probability amplitudes and $|i\rangle$ are computational basis states. The exponential memory scaling ($16 \cdot 2^n$ bytes for complex128) limits practical simulation to approximately 40--50 qubits on current hardware~\cite{QuantumSimulation2025}.
This work uses pure-state state-vector simulation; mixed states represented by density matrices are out of scope here and would require $O(4^n)$ storage.

Quantum gates are represented as unitary matrices applied to the state vector. Single-qubit gates require $O(2^n)$ operations, while two-qubit gates maintain the same asymptotic complexity but with larger constant factors due to tensor product structure. Section~\ref{sec:gate-ops} details how these operations are implemented with reshape-permute-contract kernels.

\subsection{GPU Acceleration Paradigms}

Traditional discrete GPU architectures separate host (CPU) and device (GPU) memory spaces. Data must be explicitly transferred across the PCIe bus before computation and results copied back. For quantum simulation, this creates a fundamental tension: state vectors are large (gigabytes), but individual gate operations may be compute-light, making the transfer-to-compute ratio unfavorable for shallow circuits.

GPU-accelerated quantum simulators have adopted several mitigation strategies:
\begin{itemize}
    \item \textbf{Gate batching}: Accumulating operations to amortize transfer overhead~\cite{cuQuantum2023}
    \item \textbf{Kernel fusion}: Combining sequential gates into single GPU kernels~\cite{Qsim}
    \item \textbf{Hybrid execution}: Scheduling small operations on CPU, large on GPU~\cite{QuEST}
\end{itemize}

These approaches reduce host--device movement and usually keep the state vector resident on the device during circuit execution. The remaining architectural distinction is therefore not a per-gate transfer penalty, but the cost and complexity of explicit setup/read-out copies and host-side inspection of intermediate state.

\subsection{Unified Memory Architecture}

Apple Silicon processors implement a system-on-chip design where CPU cores, GPU cores, and memory controllers share unified physical memory~\cite{UnifiedMemory2024}. Key characteristics include:
\begin{itemize}
    \item \textbf{Zero-copy access}: Both CPU and GPU access data at physical addresses without transfer
    \item \textbf{Coherent caching}: Hardware maintains cache coherency across processing elements
    \item \textbf{Scalable capacity}: Consumer systems provide 32--512\,GB unified memory
    \item \textbf{High bandwidth}: M4 Max achieves 546\,GB/s memory bandwidth~\cite{Apple2024M4}
\end{itemize}

This architecture changes the programming and data-movement model for quantum simulation: setup/read-out copies are avoided and CPU/GPU components can share arrays without explicit transfer. It does not change the exponential memory law of exact state-vector simulation or guarantee faster throughput than discrete GPUs with higher-bandwidth device memory.

\subsection{MLX Framework}

MLX~\cite{MLXFramework2024} is Apple's array computing framework optimized for Apple Silicon. Relevant features for quantum simulation include:
\begin{itemize}
    \item \textbf{NumPy-compatible API}: Familiar array semantics with broadcasting
    \item \textbf{Native complex numbers}: First-class \texttt{complex64} support
    \item \textbf{Lazy evaluation}: Deferred execution enables graph optimization
    \item \textbf{JIT compilation}: Automatic Metal shader generation
    \item \textbf{Automatic differentiation}: Built-in gradient computation
\end{itemize}

\section{System Architecture}\label{sec:architecture}

\subsection{Design Principles}

\mlxquantum follows several design principles aligned with quantum system software best practices:

\textbf{Python-first interface}: Users interact through high-level Python APIs, with MLX handling GPU acceleration transparently. No Metal shader programming is required.

\textbf{Framework compatibility}: Circuit construction follows conventions established by PennyLane and Qiskit~\cite{PennyLane2022,Qiskit2023}, enabling code migration and comparison.

\textbf{Reproducibility}: Benchmark runs produce CSV/JSON artifacts with fixed seeds, raw measured repetitions, timing summaries, and a machine-readable run manifest.

\textbf{Modularity}: Clear separation between state management, gate operations, algorithms, and benchmarking enables independent optimization of each component.

\subsection{Module Structure}

\begin{figure}[t]
\centering
\begin{tikzpicture}[
    box/.style={draw, rectangle, minimum width=2.5cm, minimum height=0.8cm, align=center, font=\footnotesize},
    arrow/.style={->, thick}
]
\node[box, fill=black!6] (state) at (0,3) {State Management\\(\texttt{mlxq.state})};
\node[box, fill=black!12] (gates) at (0,2) {Gate Operations\\(\texttt{mlxq.gates})};
\node[box, fill=black!18] (algo) at (0,1) {Algorithms\\(\texttt{mlxq.algorithms})};
\node[box, fill=black!24] (bench) at (0,0) {Benchmarking\\(\texttt{mlxq.benchmark})};
\node[box, fill=black!10] (mlx) at (3.5,1.5) {MLX\\Framework};
\node[box, fill=black!16] (metal) at (3.5,0) {Metal GPU};

\draw[arrow] (state) -- (gates);
\draw[arrow] (gates) -- (algo);
\draw[arrow] (algo) -- (bench);
\draw[arrow] (state.east) -- ++(0.5,0) |- (mlx.west);
\draw[arrow] (gates.east) -- ++(0.3,0) |- (mlx.west);
\draw[arrow] (mlx) -- (metal);
\end{tikzpicture}
\caption{Qupertino module architecture showing data flow from state management through algorithms, with MLX providing automatic Metal GPU acceleration.}
\label{fig:architecture}
\end{figure}
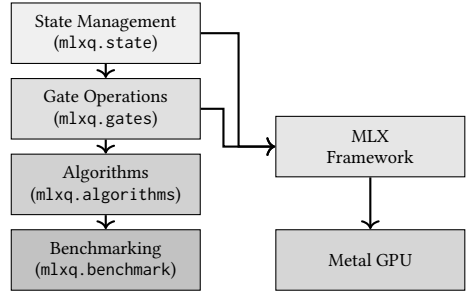

The framework comprises four principal modules illustrated in Fig.~\ref{fig:architecture}.

\subsubsection{State Management}

Quantum states are represented as MLX arrays with unified memory residency:

\begin{lstlisting}[language=Python,caption={State vector initialization}]
import mlx.core as mx
from mlxq import QuantumState

# Initialize n-qubit |0...0> state
state = QuantumState(n_qubits=10)
# state.amplitudes: mx.array, shape (1024,)
\end{lstlisting}

The \texttt{QuantumState} class provides initialization to computational basis states, normalization and probability computation, measurement sampling (exact and shot-based), and expectation value calculation for Pauli observables.

\subsubsection{Gate Operations}\label{sec:gate-ops}

Gate application uses tensor reshaping to isolate target qubits. The state vector is viewed as a rank-$n$ tensor with one index per qubit, then permuted so target indices are contiguous, contracted with the gate matrix, and permuted back. This reshape-permute-contract pattern matches standard Einstein-summation style tensor contraction workflows~\cite{SmithGray2018}. It avoids constructing full $2^n \times 2^n$ gate matrices and keeps memory traffic localized to the affected tensor slices. For a single-qubit gate $U$ on qubit $k$ of an $n$-qubit state:
\begin{enumerate}
    \item Reshape: $(2^n,) \rightarrow (2^k, 2, 2^{n-k-1})$
    \item Contract: $U$ with axis 1 (batch matmul)
    \item Flatten: $(2^k, 2, 2^{n-k-1}) \rightarrow (2^n,)$
\end{enumerate}

Two-qubit controlled gates follow analogous tensor permutation patterns. The gate library includes Pauli gates (X, Y, Z, I), Clifford gates (H, S, T, CNOT, CZ, SWAP), rotation gates (RX, RY, RZ), parameterized gates (U1, U2, U3), and multi-qubit gates (Toffoli, Fredkin).

\subsubsection{Algorithm Implementations}

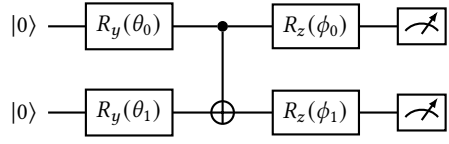
\begin{figure}[t]
\centering
\begin{quantikz}
\lstick{$\ket{0}$} & \gate{R_y(\theta_0)} & \ctrl{1} & \gate{R_z(\phi_0)} & \meter{} \\
\lstick{$\ket{0}$} & \gate{R_y(\theta_1)} & \targ{} & \gate{R_z(\phi_1)} & \meter{}
\end{quantikz}
\caption{Two-qubit hardware-efficient variational ansatz with parameterized rotations and entangling CNOT.}
\label{fig:vqe-circuit}
\end{figure}

We implement canonical quantum algorithms:

\textbf{Hardware-efficient variational ansatz optimization}: The library contains parameterized ansatze with parameter-shift gradients and the Adam optimizer, following the circuit-optimization structure used in variational algorithms~\cite{peruzzo2014variational}. In this manuscript the benchmark is a generic ansatz-optimization workload, not a molecular VQE result: no fermion-to-qubit mapping, active-space construction, or molecular Hamiltonian convergence is claimed. Fig.~\ref{fig:vqe-circuit} shows the representative two-qubit ansatz.

\textbf{Quantum Approximate Optimization Algorithm (QAOA)}: The library contains QAOA-style circuit primitives for ring MaxCut workloads~\cite{QAOA2014}. The benchmarked schedule is a fixed-depth ring phase/mixer circuit rather than a full optimizer loop. For an $n$-vertex ring with periodic boundary conditions, the reference MaxCut cost Hamiltonian is
\begin{equation}
H_C = \frac{1}{2}\sum_{i=0}^{n-1}\left(I - Z_i Z_{(i+1)\bmod n}\right),
\quad
H_M = \sum_{i=0}^{n-1} X_i .
\end{equation}
The corresponding QAOA ansatz is
\begin{equation}
|\gamma, \beta\rangle = \prod_{p=1}^{P} e^{-i\beta_p H_M} e^{-i\gamma_p H_C} |+\rangle^{\otimes n}
\end{equation}
The timing benchmark uses the same ring topology and six fixed phase/mixer layers with $\gamma_\ell = 0.6 + 0.1\ell$ and $\beta_\ell = 0.4 + 0.05\ell$, and reports circuit-execution time only. It is therefore a QAOA-style workload proxy, not evidence of MaxCut objective convergence.

\textbf{Quantum Circuit Born Machine (QCBM)}: Trains parameterized circuits via Maximum Mean Discrepancy loss to match target probability distributions.

\textbf{Quantum Fourier Transform (QFT)}: Implements the complete controlled-phase ladder~\cite{Coppersmith1994,NielsenChuang2010}:
\begin{equation}
\text{QFT}|j\rangle = \frac{1}{\sqrt{N}}\sum_{k=0}^{N-1} e^{2\pi ijk/N}|k\rangle
\end{equation}

\textbf{Hamiltonian Simulation}: Trotter-Suzuki decomposition for time evolution~\cite{Suzuki1976,Childs2019}. The main benchmark uses a transverse-field Ising schedule with open boundary conditions,
\begin{equation}
H_{\mathrm{TFIM}} = -J\sum_{i=0}^{n-2} Z_i Z_{i+1} + h\sum_{i=0}^{n-1} X_i,
\quad J=1,\quad h=0.5,\quad t=1 ,
\end{equation}
and first-order Lie-Trotter steps
\begin{equation}
e^{-iHt} \approx
\left[
\prod_{i=0}^{n-2} e^{iJ Z_i Z_{i+1}\Delta t}
\prod_{i=0}^{n-1} e^{-ihX_i\Delta t}
\right]^r,\quad \Delta t=t/r .
\end{equation}
The framework also exposes Heisenberg/XXZ-style schedules
$H_{\mathrm{XXZ}}=J\sum_i(X_iX_{i+1}+Y_iY_{i+1}+\Delta Z_iZ_{i+1})$
and random-field variants, but the timing table below reports the TFIM-style product-formula benchmark only. A small-system Trotter-error curve against exact diagonalization is reported in Table~\ref{tab:trotter-error}.

\begin{figure}[t]
\centering
\begin{quantikz}
\lstick{$\ket{0}$} & \gate{H} & \ctrl{1} & \gate{H} & \gate{X} & \ctrl{1} & \gate{X} & \gate{H} & \meter{} \\
\lstick{$\ket{0}$} & \gate{H} & \gate{Z} & \gate{H} & \gate{X} & \gate{Z} & \gate{X} & \gate{H} & \meter{}
\end{quantikz}
\caption{Two-qubit Grover search circuit with oracle (CZ) marking $|11\rangle$ and a standard diffusion operator.}
\label{fig:grover-circuit}
\end{figure}
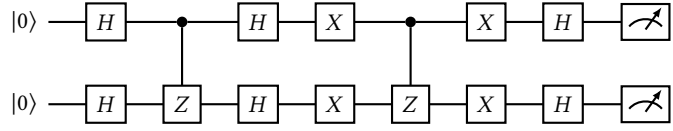

\textbf{Grover Search}: Amplitude amplification with diffusion operator for quadratic speedup over classical search~\cite{grover1996fast}. The oracle marks target states, while the diffusion operator $D = 2|\psi\rangle\langle\psi| - I$ amplifies marked amplitudes. In the two-qubit example of Fig.~\ref{fig:grover-circuit}, the marked state is $|11\rangle$.

\textbf{GHZ State Preparation}: Greenberger-Horne-Zeilinger states $|\text{GHZ}_n\rangle = \frac{1}{\sqrt{2}}(|0\rangle^{\otimes n} + |1\rangle^{\otimes n})$ are prepared with a Hadamard gate followed by $n-1$ CNOT gates, achieving maximal multipartite entanglement in linear circuit depth.

\subsection{OpenQASM Integration}

We implement an OpenQASM 2.0 parser supporting standard gate sets, parameterized rotations, user-defined gate inlining with nested definitions, and register-wide operations~\cite{OpenQASM2017}. This enables direct execution of QASMBench~\cite{QASMBench2020} circuits for standardized benchmarking.

Table~\ref{tab:qasm-coverage} summarizes OpenQASM coverage.

\begin{table}[t]
\caption{OpenQASM 2.0 gate coverage}
\label{tab:qasm-coverage}
\centering
\footnotesize
\begin{tabular}{ll}
\toprule
Category & Supported Gates \\
\midrule
Single-qubit & X, Y, Z, H, S, S$^\dagger$, T, T$^\dagger$, SX \\
Rotations & RX, RY, RZ, U1, U2, U3 \\
Two-qubit & CX, CZ, SWAP, iSWAP, CH, CRX/CRY/CRZ \\
Three-qubit & CCX (Toffoli), CSWAP (Fredkin) \\
\bottomrule
\end{tabular}
\end{table}

\section{Experimental Evaluation}\label{sec:evaluation}

\subsection{Experimental Setup}

\textbf{Hardware}: Apple M1 Max SoC with 10-core CPU (8 performance + 2 efficiency), 32-core GPU, 32\,GB unified memory, 400\,GB/s memory bandwidth, running macOS 26.3 (arm64).

\textbf{Software}: Python 3.11.3, MLX 0.30.6, complex64 state vectors. Active cooling enabled; measurements on AC power with display off.

\textbf{Protocol}: Benchmark runs separate warmup executions from measured repetitions. The current harness exposes \texttt{--warmups} and \texttt{--repeats} in the command-line runners and the QuantumStudio UI; summary timing uses the mean of measured repetitions only. Each run writes legacy plotting files, raw-run CSV/JSON files, timing summaries, and a shared manifest. The full-suite manifest records the git commit, Python/MLX/package versions, platform information, benchmark environment variables, and synchronization method; the July~4 four-way, Kokkos, shader-sweep, and QASM-sweep manifests each carry a provenance block recording the git commit, package versions, command line, environment flags, and synchronization sequence, alongside the platform, warmups/repeats, interleaving order, and quietness probe. The timing window closes only after a fixed synchronization sequence executes in order: a device-side \texttt{mx.eval} of the full probability vector ($|\text{amplitude}|^2$), then \texttt{mx.eval} of a scalar probe, then \texttt{mx.metal.synchronize} (present in MLX 0.30.6, the pinned version); only after all three return is the wall clock read. No host-side list conversion occurs inside the timing window. The comparison-table evidence run uses two warmups and ten measured repeats per cell for \mlxquantum, PennyLane, and Qiskit Aer, with the three backends interleaved round-robin within every repeat; reported uncertainty is a t-based 95\% CI computed as $t_{0.975,9}\,s/\sqrt{n} = 2.262\,s/\sqrt{10}$. Raw per-repeat values are published for every row, and warmup adequacy is verifiable from them (in the earlier three-repeat sweep, QFT at 25 qubits measured 792\,ms on the warmup and then 683/688/694\,ms, showing that kernel compilation is amortized before the window opens). Wall-clock varies with unified-memory pressure and concurrent GPU load between sessions, up to $\sim$2$\times$ on some workloads, which is exactly why the comparison campaign interleaves backends and records a quietness probe in its manifest. The scaling figures, the snapshot table, and the variational table come from the full 21-family batch (\texttt{bench.sh --repro}; one warmup, five measured repeats per size, reported as mean $\pm$ std; \path{evidence_artifacts/full_suite_20260702/}), while Table~\ref{tab:diagnostic-timings} retains the earlier three-repeat sweep, whose phase-estimation row extends to 25 qubits. Absolute values differ across these runs within the session-variance envelope; the batch additionally self-heats because the heavy Hamiltonian families run back-to-back.

\textbf{External baselines}: This revision reports same-machine PennyLane and Qiskit Aer CPU baselines for six gate-identical workloads: QFT, the fixed ring-QAOA schedule, TFIM Trotter evolution, phase estimation, the Grover diffusion proxy, and GHZ preparation. Baseline circuits reproduce the \mlxquantum gate sequences exactly (including the $\theta=-2J\,\delta t$ convention mapping for \texttt{RZZ}/\texttt{IsingZZ}), and small-system output states were cross-validated against \mlxquantum. The input gate counts at 25 qubits are identical across the three tools, with 325 gates for QFT, 300 for the ring-QAOA schedule, 980 for TFIM Trotter, 348 for phase estimation, 149 for the Grover proxy, and 25 for GHZ, and the Aer circuits carry one additional non-unitary \texttt{save\_statevector} instruction. All PennyLane circuits, including QFT, are built as the explicit gate-identical ladder; an earlier revision had measured PennyLane QFT with the \texttt{qml.QFT} template, whose decomposition appends a final SWAP layer, and that deviation is eliminated in the current campaign. Qiskit circuits are transpiled at \texttt{optimization\_level=0}, a setting chosen to preserve the gate identity of the input circuits rather than to handicap Aer; Aer's runtime gate fusion, which operates at execution time inside the simulator, and its thread parallelism are both left at their defaults. As a fairness check on this choice, we also transpiled the same 25-qubit circuits at \texttt{optimization\_level=3}, Aer's most aggressive setting, with runtime fusion enabled at both levels. The change in Aer's runtime is at most $4\%$: QFT moves from $2339$ to $2245$\,ms as the transpiler removes 10 of 326 gates, and ring-QAOA moves from $3448$ to $3502$\,ms with no gates removed, because these structured circuits are already near-minimal. The shader tier stays $\NumQftMetalVsAerOptThree\times$ and $\NumQaoaMetalVsAerOptThree\times$ faster than Aer at its best transpiler setting (\path{aer_sensitivity_20260706.json}), so the \texttt{optimization\_level=0} baseline does not materially flatter our result. Each backend therefore applies its own kernel-level execution optimizations to the same input gate sequence; this is the comparison we report. The comparison campaign uses two warmups and ten interleaved measured repeats per cell, records raw runs and a session manifest with a quietness probe, and forces state-vector materialization in every backend. Qiskit 2.4.2, qiskit-aer 0.17.2, PennyLane 0.43.0, and MLX 0.30.6 were installed in the project runtime environment. One backend-configuration asymmetry deserves disclosure: Aer's CPU statevector engine is multi-threaded by default, whereas the \texttt{lightning.qubit} macOS wheel links no OpenMP runtime and applies gates single-threaded. We therefore also measured PennyLane's OpenMP-parallel \texttt{lightning.kokkos} plugin (v0.45.0, macOS arm64 wheel) in a separate paired campaign (Table~\ref{tab:kokkos}): \mlxquantum remains fastest in 16 of 18 cells, by $3.0$--$8.1\times$ at 25 qubits, with every 25-qubit winning interval excluding parity; one 15-qubit cell (ring-QAOA, paired ratio $2.62\pm2.90$) has an interval that spans parity and is not counted as a significant win. Threading engagement was verified explicitly; the default Kokkos configuration matches \texttt{OMP\_NUM\_THREADS=10} to within measurement noise, and forcing a single thread costs only $\sim$20\%, indicating the workload is memory-bandwidth-bound rather than compute-bound on this host.

\begin{table}[t]
\caption{Paired campaign vs.\ PennyLane \texttt{lightning.kokkos} (OpenMP; two warmups, ten interleaved repeats; means in seconds; ratio is the paired per-repeat kokkos/\mlxquantum speedup, mean $\pm$ t-based 95\% CI). Absolute values are session-specific; the ratios are paired within repeats.}
\label{tab:kokkos}
\centering
\footnotesize
\setlength{\tabcolsep}{4pt}
\begin{tabular}{@{}lrrrr@{}}
\toprule
Workload & Qubits & \mlxquantum (s) & kokkos (s) & Paired ratio \\
\midrule
QFT & 15 & 0.0046 & 0.0063 & $1.39 \pm 0.09$ \\
QFT & 20 & 0.0274 & 0.1044 & $3.86 \pm 0.33$ \\
QFT & 25 & 0.720 & 5.807 & $8.06 \pm 0.64$ \\
Ring-QAOA & 15 & 0.0081 & 0.0208 & $2.62 \pm 2.90$ \\
Ring-QAOA & 20 & 0.0391 & 0.2215 & $5.69 \pm 0.31$ \\
Ring-QAOA & 25 & 2.561 & 13.577 & $5.63 \pm 1.26$ \\
TFIM Trotter & 15 & 0.0218 & 0.0327 & $1.53 \pm 0.17$ \\
TFIM Trotter & 20 & 0.1180 & 1.0267 & $8.76 \pm 0.56$ \\
TFIM Trotter & 25 & 6.864 & 40.201 & $6.17 \pm 1.08$ \\
Phase estimation & 15 & 0.0052 & 0.0065 & $1.24 \pm 0.04$ \\
Phase estimation & 20 & 0.0254 & 0.1181 & $4.69 \pm 0.37$ \\
Phase estimation & 25 & 0.868 & 5.904 & $6.80 \pm 0.17$ \\
Grover proxy & 15 & 0.0177 & 0.0050 & $0.34 \pm 0.11$ \\
Grover proxy & 20 & 0.0301 & 0.1213 & $4.04 \pm 0.33$ \\
Grover proxy & 25 & 1.045 & 4.178 & $4.00 \pm 0.10$ \\
GHZ & 15 & 0.0014 & 0.0011 & $0.74 \pm 0.04$ \\
GHZ & 20 & 0.0068 & 0.0117 & $1.78 \pm 0.23$ \\
GHZ & 25 & 0.1517 & 0.4457 & $3.00 \pm 0.26$ \\
\bottomrule
\multicolumn{5}{@{}l@{}}{\footnotesize Artifacts: \path{evidence_artifacts/kokkos_campaign_20260704/} (raw per-repeat CSV, manifest, quietness probe 0.645\,s).}\\
\end{tabular}
\end{table} Raw artifacts are under \path{evidence_artifacts/interleaved_4way_20260704/} (four-way comparison campaign) and \path{evidence_artifacts/mlx_repro_20260702/} (scaling sweep).

\textbf{Kernel-dispatch ablation}: To separate the structured-dispatch speedup from MLX GPU execution itself, the backend exposes \texttt{MLXQ\_DENSE\_ONLY=1}, which routes every gate through the generic dense reshape/transpose/matmul path while keeping MLX, the device, and the protocol fixed. In a paired run at 25 qubits in the same session as the four-way campaign, the dense-only path takes $34.7$\,s for QFT and $41.8$\,s for ring-QAOA versus $1.07$\,s and $1.68$\,s with dispatch enabled. The resulting $32.5\times$ and $24.8\times$ ratios are attributable to kernel specialization alone rather than to GPU execution as such (\path{evidence_artifacts/interleaved_4way_20260704/kernel_dispatch_ablation_idle.json}). Correctness parity between the two paths is regression-tested (Table~\ref{tab:parity}).

\subsection{Benchmark Workloads, Depth, and Snapshot}

The benchmark families were chosen so that each workload has a recognizable counterpart in an established suite, which lets readers compare our measurements against results published for PennyLane, Yao.jl, Qulacs, QuantumToolbox.jl, and QASMBench without translating between circuit conventions; Table~\ref{tab:coverage} lists this alignment. Because runtime depends on circuit structure as much as on qubit count, we also report circuit depth for each family, counted as the number of sequential gate stages, where parallel single-qubit layers on disjoint wires contribute one stage and serialized two-qubit cascades or repeated ansatz blocks contribute one stage per operation that cannot be parallelized. Table~\ref{tab:benchmark-depth} gives the resulting depth expressions for representative families. With these conventions fixed, Table~\ref{tab:scaling-full} presents the current 25-qubit snapshot, taken from the full 21-family reproducibility batch (\texttt{bench.sh --repro}; one warmup and five measured repeats per size, structured-gate kernel dispatch enabled; timing summaries under \path{evidence_artifacts/full_suite_20260702/}). The same-machine external comparison for the six gate-identical workloads follows in Sec.~\ref{sec:pennylane-baseline}, and additional 25-qubit workloads from the earlier scaling sweep appear in Table~\ref{tab:diagnostic-timings}.

\begin{table}[t]
\caption{Benchmark coverage by framework alignment}
\label{tab:coverage}
\centering
\begin{tabular}{ll}
\toprule
Framework & Workloads \\
\midrule
PennyLane~\cite{PennyLane2022} & VQE, QAOA, variational circuits, QCBM \\
Yao.jl~\cite{YaoJL2024} & Random circuits, QFT, time evolution, Trotter \\
Qulacs~\cite{Qulacs} & QFT, Grover, random circuits \\
QuantumToolbox.jl~\cite{QuantumToolboxJulia2024} & Hamiltonian sim., steady state \\
QASMBench~\cite{QASMBench2020} & OpenQASM circuits (small/medium scale) \\
\bottomrule
\end{tabular}
\end{table}

\begin{table}[t]
\caption{Representative benchmark depth formulas}
\label{tab:benchmark-depth}
\centering
\footnotesize
\begin{tabular}{lcc}
\toprule
Benchmark & Depth Expression & Example Depth \\
\midrule
GHZ ($n$ qubits) & $1 + (n-1)$ & $n=25 \Rightarrow 25$ \\
QCBM ($n$ qubits, $L$ layers) & $L(2 + n-1)$ & $n=25, L=9 \Rightarrow 234$ \\
QFT ($n$ qubits) & $n + \frac{n(n-1)}{2} + \lfloor n/2 \rfloor$ & $n=25 \Rightarrow 337$ \\
HE ansatz optimization (2 qubits) & 3 staged layers & 3 \\
\bottomrule
\multicolumn{3}{@{}l@{}}{\footnotesize The QFT expression follows the textbook circuit including the final $\lfloor n/2 \rfloor$ swap}\\
\multicolumn{3}{@{}l@{}}{\footnotesize stages; the measured QFT workload omits the swap layer (325 gates at $n{=}25$).}\\
\end{tabular}
\end{table}

\begin{table}[t]
\caption{Current benchmark snapshot on Apple M1 Max (mean $\pm$ std; full-suite batch via \texttt{bench.sh --repro}, one warmup, five measured repeats)}
\label{tab:scaling-full}
\centering
\begin{tabular}{lrr}
\toprule
Algorithm & Qubits & Time (s) \\
\midrule
QFT & 25 & $0.822 \pm 0.081$ \\
QAOA-style ring schedule & 25 & $2.703 \pm 0.683$ \\
QCBM (9 layers) & 25 & $2.345 \pm 0.628$ \\
TFIM Hamiltonian evolution ($r{=}20$) & 25 & $5.123 \pm 0.518$ \\
Time Evolution ($r{=}12$) & 25 & $3.931 \pm 0.227$ \\
\bottomrule
\multicolumn{3}{l}{\footnotesize Same-machine external comparison for gate-identical workloads: Table~\ref{tab:pennylane-baseline}.}
\end{tabular}
\end{table}

\begin{table}[t]
\caption{Additional 25-qubit workloads (mean $\pm$ std, three measured repeats, single sweep)}
\label{tab:diagnostic-timings}
\centering
\footnotesize
\begin{tabular}{lrrl}
\toprule
Algorithm & Qubits & Time (s) & Status \\
\midrule
Phase Estimation & 25 & $2.19 \pm 0.11$ & Measured \\
Random Circuit (depth 6) & 25 & $1.98 \pm 0.04$ & Measured \\
Variational Circuit (4 layers) & 25 & $2.28 \pm 0.03$ & Measured \\
Grover Search (proxy, 1 iter.) & 25 & $1.03 \pm 0.01$ & Measured \\
GHZ State & 25 & $0.24 \pm 0.02$ & Measured \\
\bottomrule
\multicolumn{4}{@{}l@{}}{\footnotesize Phase estimation, Grover, and GHZ also appear in Table~\ref{tab:pennylane-baseline}, which uses the}\\
\multicolumn{4}{@{}l@{}}{\footnotesize separate ten-repeat interleaved campaign. Absolute times are campaign-specific and must}\\
\multicolumn{4}{@{}l@{}}{\footnotesize not be compared across sources (phase estimation spans $0.9$--$2.2$\,s at 25 qubits across}\\
\multicolumn{4}{@{}l@{}}{\footnotesize same-day sessions; Sec.~\ref{sec:pennylane-baseline} quantifies this envelope per workload).}\\
\end{tabular}
\end{table}

\FloatBarrier

\subsection{Same-Machine External Baselines}\label{sec:pennylane-baseline}

Table~\ref{tab:pennylane-baseline} reports same-machine timings for four backends on six gate-identical workloads: QFT, the fixed ring-QAOA schedule, TFIM Trotter evolution (20 first-order steps, $J{=}1$, $h{=}0.5$, $t{=}1$), phase estimation, the Grover diffusion proxy, and GHZ preparation. The two \mlxquantum columns are the two tiers of this paper: the pure-MLX tier dispatches diagonal gates to broadcast phase multiplies, controlled single-qubit gates to half-state updates, SWAP to an axis permutation, and fuses runs of equal-angle ZZ terms into one cached diagonal multiply; the Metal tier additionally routes fusable layers to the hand-written kernels of Section~\ref{sec:shaders} (\texttt{MLXQ\_METAL\_KERNELS=1}). Both receive exactly the gate stream that Aer and PennyLane receive; small-system states were validated against dense execution and against both external baselines ($\le 2\times10^{-6}$ max amplitude deviation at $n{=}4$--$7$). All rows come from a single interleaved campaign (\texttt{tools/interleaved\_campaign.py}): for every workload and size, the four backends execute round-robin inside each of ten measured repeats, after two warmups per backend and a machine-quietness probe, so session-level drift and unified-memory pressure affect all four backends equally. The Metal tier is fastest by mean runtime in ALL 18 comparison cells, at 25 qubits by $23$--$47\times$ over Aer and $19$--$95\times$ over PennyLane (paired ratios below). The GPU-backed pure-MLX tier is fastest among the same-host CPU trio on all six 25-qubit workloads by mean --- five of six by the stricter paired-ratio test, since the Grover proxy at 25 qubits is a statistical tie with Aer (paired ratio $1.09\pm0.14$); the smallest gate-sparse circuits at 15 qubits (GHZ, Grover) still favor the CPU baselines' lower per-gate launch overhead. Baseline absolute times are somewhat higher here than in solo runs because interleaving exposes every backend to the same memory pressure that \mlxquantum's 256\,MB working set generates; that is the intended control, not an artifact.

Because the backends run interleaved within every repeat, the raw data also supports paired statistics, which are stronger than independent interval comparisons. Computing the per-repeat speedup ratio inside each of the ten repeats and applying the same t-based interval yields, at 25 qubits for the Metal tier: QFT $\NumQftAerRatio\pm\NumQftAerCI$ (Aer/Metal) and $\NumQftPlRatio\pm\NumQftPlCI$ (PennyLane/Metal); ring-QAOA $\NumQaoaAerRatio\pm\NumQaoaAerCI$ and $\NumQaoaPlRatio\pm\NumQaoaPlCI$; TFIM $\NumTfimAerRatio\pm\NumTfimAerCI$ and $\NumTfimPlRatio\pm\NumTfimPlCI$; phase estimation $\NumPhaseEstAerRatio\pm\NumPhaseEstAerCI$ and $\NumPhaseEstPlRatio\pm\NumPhaseEstPlCI$; Grover proxy $\NumGroverAerRatio\pm\NumGroverAerCI$ and $\NumGroverPlRatio\pm\NumGroverPlCI$; GHZ $\NumGhzAerRatio\pm\NumGhzAerCI$ and $\NumGhzPlRatio\pm\NumGhzPlCI$. For the pure-MLX tier the same intervals are: QFT $3.91\pm0.52$ and $7.84\pm0.54$; ring-QAOA $3.02\pm0.93$ and $3.72\pm0.85$; TFIM $3.17\pm0.51$ and $5.85\pm0.79$; phase estimation $4.44\pm0.57$ and $6.64\pm0.11$; Grover proxy $1.09\pm0.14$ (parity inside the interval; the one cell the pure tier does not win) and $2.45\pm0.10$; GHZ $2.53\pm0.32$ and $1.54\pm0.04$. Every Metal-tier paired interval excludes parity by a wide margin, as does the tier-versus-tier comparison (pure-MLX/Metal $9.8$--$21.3\times$ at 25 qubits). Two honesty notes on comparison with the July~2 three-way campaign (\path{evidence_artifacts/interleaved_campaign_20260702/}): the pure tier's TFIM mean improved from $9.88$ to $5.82$\,s because runtime ZZ-layer fusion entered the pure path between the campaigns, and the Aer Grover and TFIM means shifted by up to $1.4\times$ between sessions, which is why we base every comparative claim on within-session paired ratios rather than cross-campaign values. The separately-labeled QFT (FFT primitive) row times \texttt{mx.fft}-based full-register QFT with bit-reversal; the Metal tier's gate-stream QFT ($0.0591\pm0.0029$\,s) is faster than that primitive because each detected stage runs as one fused butterfly-plus-ladder pass, and the helper-level radix-4 path (two stages per launch) reaches $21.1$\,ms (Table~\ref{tab:shader}). Raw artifacts are under \path{evidence_artifacts/interleaved_4way_20260704/}.

\begin{table}[t]
\caption{Four-way same-machine comparison (mean $\pm$ t-based 95\% CI ($2.262\,s/\sqrt{10}$), seconds; two warmups, ten measured repeats, all four backends interleaved round-robin within every repeat; complex64, single M1 Max host, CPU baselines only)}
\label{tab:pennylane-baseline}
\centering
\footnotesize
\scriptsize
\setlength{\tabcolsep}{2.5pt}
\begin{tabular}{@{}llcccc@{}}
\toprule
Workload & Qubits & \mlxquantum (MLX) & \mlxquantum (Metal) & PennyLane & Aer CPU \\
\midrule
QFT & 15 & $0.00545 \pm 0.00013$ & $0.00093 \pm 0.00010$ & $0.00884 \pm 0.00098$ & $0.0107 \pm 0.0028$ \\
QFT & 20 & $0.0275 \pm 0.0026$ & $0.00259 \pm 0.00015$ & $0.1244 \pm 0.0063$ & $0.1005 \pm 0.0060$ \\
QFT & 25 & $0.7212 \pm 0.0498$ & $0.0591 \pm 0.0029$ & $5.611 \pm 0.182$ & $2.796 \pm 0.342$ \\
Ring-QAOA schedule & 15 & $0.00897 \pm 0.00111$ & $0.00176 \pm 0.00017$ & $0.00946 \pm 0.00016$ & $0.0138 \pm 0.0093$ \\
Ring-QAOA schedule & 20 & $0.0461 \pm 0.0013$ & $0.00472 \pm 0.00012$ & $0.1642 \pm 0.0008$ & $0.1273 \pm 0.0057$ \\
Ring-QAOA schedule & 25 & $2.070 \pm 0.583$ & $0.1503 \pm 0.0215$ & $6.843 \pm 0.283$ & $5.351 \pm 0.655$ \\
TFIM Trotter (20 steps) & 15 & $0.0183 \pm 0.0011$ & $0.00401 \pm 0.00037$ & $0.0389 \pm 0.0221$ & $0.0336 \pm 0.0066$ \\
TFIM Trotter (20 steps) & 20 & $0.1531 \pm 0.0136$ & $0.0135 \pm 0.0002$ & $0.9187 \pm 0.0312$ & $0.4778 \pm 0.0275$ \\
TFIM Trotter (20 steps) & 25 & $5.824 \pm 0.810$ & $0.4949 \pm 0.0348$ & $32.950 \pm 0.323$ & $17.792 \pm 0.910$ \\
Phase estimation & 15 & $0.00642 \pm 0.00070$ & $0.00171 \pm 0.00059$ & $0.00722 \pm 0.00007$ & $0.00726 \pm 0.00092$ \\
Phase estimation & 20 & $0.0463 \pm 0.0208$ & $0.0150 \pm 0.0197$ & $0.1392 \pm 0.0117$ & $0.1136 \pm 0.0167$ \\
Phase estimation & 25 & $0.9117 \pm 0.0106$ & $0.1050 \pm 0.0399$ & $6.056 \pm 0.099$ & $4.051 \pm 0.526$ \\
Grover proxy (1 iter.) & 15 & $0.00559 \pm 0.00052$ & $0.00163 \pm 0.00006$ & $0.00359 \pm 0.00008$ & $0.00274 \pm 0.00014$ \\
Grover proxy (1 iter.) & 20 & $0.0320 \pm 0.0011$ & $0.00371 \pm 0.00048$ & $0.0777 \pm 0.0238$ & $0.0215 \pm 0.0008$ \\
Grover proxy (1 iter.) & 25 & $1.112 \pm 0.014$ & $0.0521 \pm 0.0007$ & $2.725 \pm 0.128$ & $1.209 \pm 0.161$ \\
GHZ & 15 & $0.00184 \pm 0.00034$ & $0.00064 \pm 0.00022$ & $0.00133 \pm 0.00010$ & $0.00141 \pm 0.00007$ \\
GHZ & 20 & $0.0454 \pm 0.0240$ & $0.00609 \pm 0.00173$ & $0.0117 \pm 0.0002$ & $0.0288 \pm 0.0070$ \\
GHZ & 25 & $0.2726 \pm 0.0052$ & $0.0223 \pm 0.0010$ & $0.4199 \pm 0.0122$ & $0.6915 \pm 0.0947$ \\
\midrule
QFT (FFT primitive)$^{\dagger}$ & 15 & $0.00068 \pm 0.00015$ & -- & -- & -- \\
QFT (FFT primitive)$^{\dagger}$ & 20 & $0.00196 \pm 0.00004$ & -- & -- & -- \\
QFT (FFT primitive)$^{\dagger}$ & 25 & $0.0772 \pm 0.0029$ & -- & -- & -- \\
\bottomrule
\multicolumn{6}{@{}l@{}}{\footnotesize $^{\dagger}$\texttt{mx.fft}-based algorithmic primitive with bit-reversal, not gate-by-gate circuit execution (July 2 campaign);}\\
\multicolumn{6}{@{}l@{}}{\footnotesize \phantom{$^{\dagger}$}comparable only to libraries exposing a native QFT/FFT operation.}\\
\end{tabular}
\end{table}

\FloatBarrier

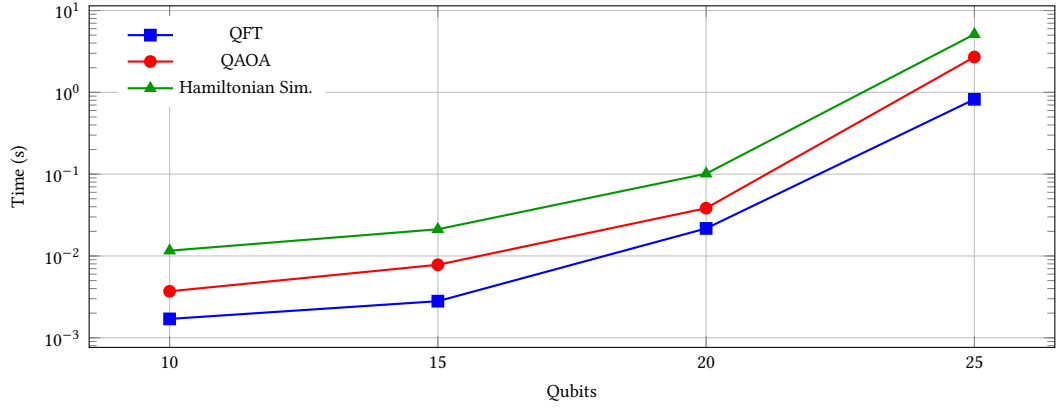
\begin{figure}[t]
\centering
\begin{tikzpicture}
\begin{axis}[
    width=0.95\columnwidth,
    height=6.1cm,
    xlabel={Qubits},
    ylabel={Time (s)},
    ymode=log,
    xtick={10,15,20,25},
    grid=major,
    legend pos=north west,
    legend style={font=\footnotesize, fill=white, draw=none},
    tick label style={font=\footnotesize},
    label style={font=\footnotesize}
]
\addplot[blue,mark=square*,thick] coordinates {(10,0.0017) (15,0.0028) (20,0.0217) (25,0.822)};
\addlegendentry{QFT}
\addplot[red,mark=*,thick] coordinates {(10,0.0037) (15,0.0078) (20,0.0384) (25,2.703)};
\addlegendentry{QAOA}
\addplot[green!60!black,mark=triangle*,thick] coordinates {(10,0.0116) (15,0.0212) (20,0.1013) (25,5.123)};
\addlegendentry{Hamiltonian Sim.}
\end{axis}
\end{tikzpicture}
\caption{Measured scaling curves (Apple M1 Max, full-suite batch, mean of five repeats). All three workloads show monotonic growth with qubit count; the same-machine external comparison uses Table~\ref{tab:pennylane-baseline}.}
\label{fig:asset-benchmarks}
\end{figure}

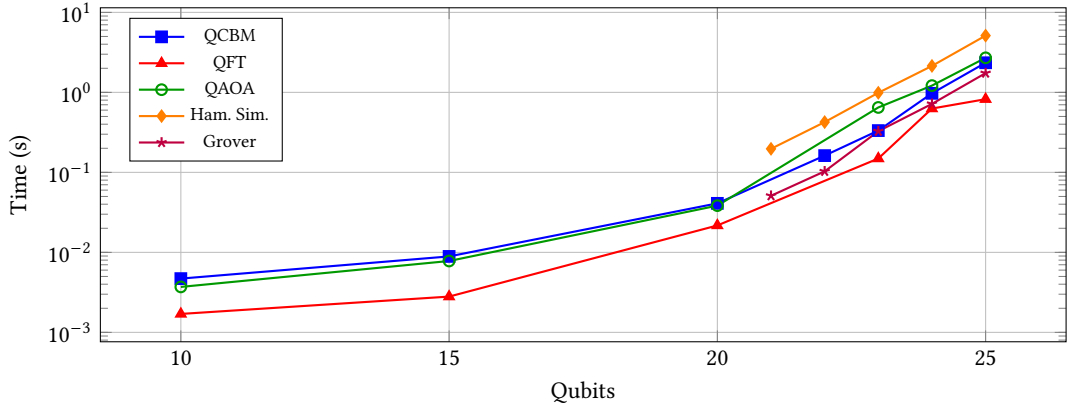
\begin{figure}[t]
\centering
\begin{tikzpicture}
\begin{axis}[
    width=0.95\columnwidth,
    height=6cm,
    ymode=log,
    xlabel={Qubits},
    ylabel={Time (s)},
    legend pos=north west,
    legend style={font=\footnotesize},
    grid=major,
    xtick={10,15,20,25},
]
\addplot[blue,mark=square*,thick] coordinates {
    (10,0.0047) (15,0.0089) (20,0.0410) (22,0.162) (23,0.333) (24,0.973) (25,2.345)
};
\addlegendentry{QCBM}
\addplot[red,mark=triangle*,thick] coordinates {
    (10,0.0017) (15,0.0028) (20,0.0217) (23,0.149) (24,0.628) (25,0.822)
};
\addlegendentry{QFT}
\addplot[green!60!black,mark=o,thick] coordinates {
    (10,0.0037) (15,0.0078) (20,0.0384) (23,0.648) (24,1.215) (25,2.703)
};
\addlegendentry{QAOA}
\addplot[orange,mark=diamond*,thick] coordinates {
    (21,0.197) (22,0.425) (23,0.988) (24,2.141) (25,5.123)
};
\addlegendentry{Ham. Sim.}
\addplot[purple,mark=star,thick] coordinates {
    (21,0.051) (22,0.103) (23,0.327) (24,0.718) (25,1.740)
};
\addlegendentry{Grover}
\end{axis}
\end{tikzpicture}
\caption{Measured exponential-scaling curves for benchmark algorithms (M1 Max, 32\,GB unified memory, full-suite batch, mean of five repeats). Deep circuits confirm the expected $O(2^n)$ trend at 21--25 qubits; below that, per-kernel launch overhead flattens the curves. The same-machine external comparison uses Table~\ref{tab:pennylane-baseline}.}
\label{fig:scaling}
\end{figure}

\begin{figure}[t]
\centering
\begin{tikzpicture}
\begin{axis}[
    width=0.95\columnwidth,
    height=7.0cm,
    xlabel={Qubits},
    ylabel={Time (s)},
    ymode=log,
    xmin=10, xmax=25,
    xtick={10,15,20,25},
    grid=major,
    legend style={font=\scriptsize, at={(0.02,0.98)}, anchor=north west, fill=white, draw=none},
    tick label style={font=\footnotesize},
    label style={font=\footnotesize}
]
\addplot[blue,mark=square*,thick] coordinates {(10,0.0017) (15,0.0028) (20,0.0217) (25,0.822)};
\addlegendentry{QFT}
\addplot[red,mark=*,thick] coordinates {(10,0.0037) (15,0.0078) (20,0.0384) (25,2.703)};
\addlegendentry{QAOA}
\addplot[teal!70!black,mark=triangle*,thick] coordinates {(10,0.0047) (15,0.0089) (20,0.0410) (25,2.345)};
\addlegendentry{QCBM}
\addplot[green!60!black,mark=diamond*,thick] coordinates {(21,0.197) (22,0.425) (23,0.988) (24,2.141) (25,5.123)};
\addlegendentry{Hamiltonian Sim.}
\addplot[purple,mark=star,thick] coordinates {(21,0.051) (22,0.103) (23,0.327) (24,0.718) (25,1.740)};
\addlegendentry{Grover}
\addplot[orange!90!black,mark=o,thick] coordinates {(10,0.0077) (15,0.0152) (20,0.0687) (25,3.931)};
\addlegendentry{Time Evolution}
\end{axis}
\end{tikzpicture}
\caption{Cross-family benchmark comparison on Apple M1 Max (single-panel log-scale plot, full-suite batch). QFT remains the fastest workload at fixed qubit count; Hamiltonian-evolution families exhibit the largest runtimes near 25 qubits, and gate-sparse workloads such as Grover sit near the launch-overhead floor until 21+ qubits.}
\label{fig:all-benchmarks-comparison}
\end{figure}
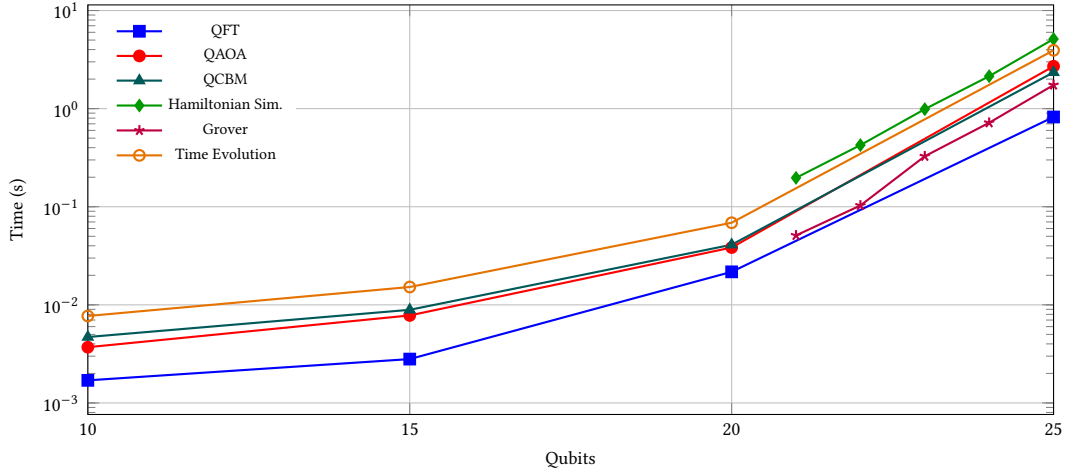

\subsection{Detailed Algorithm Performance}

The variational rows deserve a word on what they do and do not measure. QCBM uses 9-layer hardware-efficient ansatze with alternating rotation and entanglement layers, while the QAOA-style row times the fixed ring phase/mixer schedule described above, so neither includes objective evaluation or optimizer convergence; Table~\ref{tab:variational} reports the resulting circuit-execution timing for 23--25 qubits.

\begin{table}[t]
\caption{Variational circuit performance (mean $\pm$ std; full-suite batch, five measured repeats)}
\label{tab:variational}
\centering
\begin{tabular}{lrrr}
\toprule
Algorithm & 23q (s) & 24q (s) & 25q (s) \\
\midrule
QCBM & $0.333 \pm 0.002$ & $0.973 \pm 0.231$ & $2.345 \pm 0.628$ \\
QAOA & $0.648 \pm 0.737$ & $1.215 \pm 0.478$ & $2.703 \pm 0.683$ \\
\bottomrule
\end{tabular}
\end{table}

A different cost regime appears when the circuit sits inside a training loop. The hardware-efficient ansatz optimization diagnostic exhibits steep scaling because iterative optimization with parameter-shift gradients requires two forward passes per parameter per iteration, and the number of parameters grows with the ansatz width; Table~\ref{tab:vqe} shows the archived measurements.

\begin{table}[t]
\caption{Hardware-efficient ansatz scaling diagnostic (Adam, parameter-shift gradients, 100 iterations; archived run predating the structured-gate kernel dispatch)}
\label{tab:vqe}
\centering
\begin{tabular}{rrr}
\toprule
Qubits & Time (s) & Time/Iter (s) \\
\midrule
10 & 3.46 & 0.035 \\
11 & 15.44 & 0.154 \\
12 & 77.72 & 0.777 \\
13 & 373.70 & 3.74 \\
14 & 1942.59 & 19.43 \\
15 & 16718.00 & 167.18 \\
\bottomrule
\end{tabular}
\end{table}

The 15-qubit ansatz-optimization benchmark ($\sim$4.6 hours) is a single-run optimization-heavy diagnostic, not a molecular-chemistry result. The current manuscript does not report a fermion-to-qubit mapping, an active-space construction, or convergence against an exact H$_2$ ground-state energy; those are future validation tasks rather than current claims. Gradient-free methods or hardware-accelerated automatic differentiation could extend this range.

The full batch also covers the Heisenberg, XXZ, random-field, long-range Ising, and ladder evolution variants, which run in 18--33\,s at 25 qubits because each Trotter step applies three Pauli-pair sweeps (XX, YY, ZZ) rather than the TFIM's one, and the density-matrix steady-state diagnostic, which retains the generic dense path and is the slowest remaining family ($\sim$292\,s at its 12-qubit cap). Hamiltonian simulation, finally, uses Trotter-Suzuki decomposition with a configurable step count, and its runtime scales with both the step count and the number of bonds. Table~\ref{tab:hamiltonian} reports the current first-order TFIM measurement with $r=64$ product-formula steps, $J=1$, $h=0.5$, $t=1$, and open boundary conditions, taken with one warmup and three measured repeats at a gate count of $r\,(2n-1)$ per run, and Table~\ref{tab:trotter-error} gives the corresponding small-system error trend versus $r$.

\begin{table}[t]
\caption{Hamiltonian simulation scaling ($r=64$; one warmup, three measured repeats, mean $\pm$ std)}
\label{tab:hamiltonian}
\centering
\begin{tabular}{rrrr}
\toprule
Qubits & Time (s) & Steps & Gates \\
\midrule
21 & $1.38 \pm 0.04$ & 64 & 2,624 \\
22 & $2.46 \pm 0.21$ & 64 & 2,752 \\
23 & $6.08 \pm 0.20$ & 64 & 2,880 \\
24 & $14.35 \pm 0.38$ & 64 & 3,008 \\
25 & $35.39 \pm 0.75$ & 64 & 3,136 \\
\bottomrule
\end{tabular}
\end{table}

\subsection{Beyond Exact State-Vector: MPS Scale-Out}\label{sec:mps-scaleout}

Exact state-vector simulation ends near 30 qubits on 32\,GB of memory, but the same artifact ships a Matrix Product State backend (\texttt{MLXQ\_BACKEND=mps}) whose cost scales with entanglement rather than with $2^n$, and recent emulator benchmarking shows that MPS methods dominate classical simulation at utility scale. Table~\ref{tab:mps-scaleout} therefore extends the measured evidence past the exact-simulation ceiling for two limited-entanglement workloads. GHZ preparation, whose bond dimension is exactly two, runs to 150 qubits in under 60\,ms. TFIM TEBD evolution ($r{=}12$ first-order steps, bond cap $D_{\max}{=}64$) reaches 100 qubits in roughly 22\,s while the MPS tensors occupy only 1.2--5.6\,MB, four orders of magnitude below the corresponding state-vector footprint. Two caveats keep this honest. First, the TEBD rows are approximate: the bond dimension saturates the $D_{\max}{=}64$ cap at every size shown, and the harness flags truncation for all of them, so these timings describe bounded-bond evolution rather than exact dynamics. Second, run-to-run variance on the SVD-dominated path is large (standard deviations up to $\sim$40\% of the mean), reflecting the many small, shape-varying factorizations this backend performs; the MPS rows are scale-feasibility evidence, not precision benchmarks, and the structured-gate dispatch study in Sec.~\ref{sec:pennylane-baseline} remains the paper's quantitative core.

\begin{table}[t]
\caption{MPS-backend scale-out (mean $\pm$ std, one warmup, three measured repeats; $D_{\max}=64$)}
\label{tab:mps-scaleout}
\centering
\footnotesize
\begin{tabular}{lrrl}
\toprule
Workload & Qubits & Time & Notes \\
\midrule
GHZ & 30 & $11.7 \pm 1.7$\,ms & analytic bond dim.\ 2 (summary records 0; instrumentation under repair) \\
GHZ & 50 & $19.3 \pm 1.3$\,ms & exact \\
GHZ & 100 & $38.1 \pm 1.1$\,ms & exact \\
GHZ & 150 & $58.6 \pm 8.8$\,ms & exact \\
TFIM TEBD ($r{=}12$) & 30 & $3.70 \pm 2.31$\,s & truncated at $D_{\max}$; 1.2\,MB tensors \\
TFIM TEBD ($r{=}12$) & 50 & $9.88 \pm 4.34$\,s & truncated; 2.5\,MB \\
TFIM TEBD ($r{=}12$) & 75 & $21.7 \pm 7.3$\,s & truncated; 4.0\,MB \\
TFIM TEBD ($r{=}12$) & 100 & $21.6 \pm 9.4$\,s & truncated; 5.6\,MB \\
\bottomrule
\multicolumn{4}{@{}l@{}}{\footnotesize Artifacts: \path{evidence_artifacts/mps_scaleout_20260702/} (timing summaries, per-site bond CSVs).}\\
\end{tabular}
\end{table}

\subsection{Memory Efficiency}

State-vector memory scales as $8 \cdot 2^n$ bytes for complex64. Table~\ref{tab:memory-capacity} summarizes absolute memory footprint and practical capacity outlook across representative Apple Silicon configurations.

\begin{table}[t]
\caption{State-vector memory scaling and practical capacity outlook (complex64)}
\label{tab:memory-capacity}
\centering
\begin{tabular}{lrr}
\toprule
Configuration / scale & Memory & Practical qubit range \\
\midrule
20 qubits & 8 MB & comfortable on tested systems \\
25 qubits & 256 MB & fully measured in this work \\
30 qubits & 8 GB & feasible on 32 GB machines \\
35 qubits & 256 GB & requires ultra-class memory systems \\
\midrule
M1 Max (32 GB) & 32 GB unified & up to $\sim$30 qubits \\
M2 Ultra (192 GB) & 192 GB unified & up to $\sim$33 qubits \\
M3 Ultra (512 GB) & 512 GB unified & up to $\sim$35 qubits \\
\bottomrule
\end{tabular}
\end{table}

The M1 Max's 32\,GB unified memory can store a 30-qubit complex64 state vector in principle, but full algorithmic workloads also require intermediate allocations, compilation overhead, and time-to-solution headroom. The measured snapshot in this work is therefore centered on 25-qubit runs, while 30--35 qubit entries are capacity outlooks rather than validated performance claims. Higher-memory Ultra configurations improve memory headroom, but unified memory does not alter the exponential memory law; it reduces transfer overhead and software complexity for a given problem size.

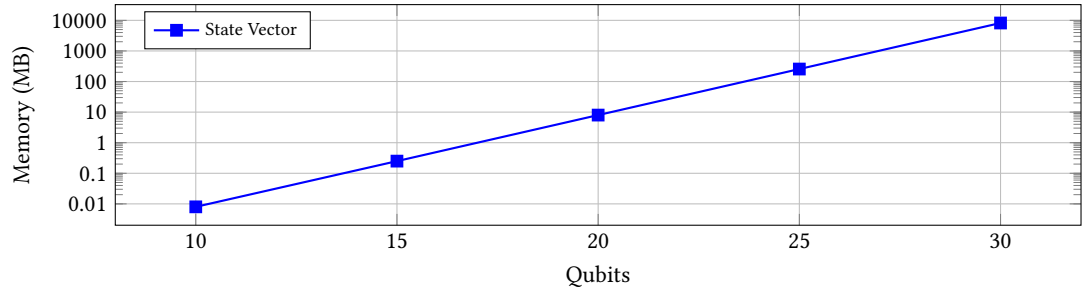
\begin{figure}[t]
\centering
\begin{tikzpicture}
\begin{axis}[
    width=0.95\columnwidth,
    height=4.5cm,
    xlabel={Qubits},
    ylabel={Memory (MB)},
    ymode=log,
    grid=major,
    xtick={10,15,20,25,30},
    ytick={0.01,0.1,1,10,100,1000,10000},
    yticklabels={0.01,0.1,1,10,100,1000,10000},
    legend pos=north west,
    legend style={font=\footnotesize},
]
\addplot[blue,mark=square*,thick] coordinates {
    (10,0.008) (15,0.25) (20,8) (25,256) (30,8192)
};
\addlegendentry{State Vector}
\end{axis}
\end{tikzpicture}
\caption{Memory scaling with qubit count (complex64). A 30-qubit state occupies 8\,GB; 35 qubits requires 256\,GB, approaching M4 Ultra-class capacity.}
\label{fig:memory-scaling}
\end{figure}

Fig.~\ref{fig:memory-scaling} shows the memory-time relationship across qubit scales.

\subsection{Unified Memory Advantage}

\begin{figure}[t]
\centering
\begin{tikzpicture}[scale=0.8]
% Discrete GPU
\node[draw, rectangle, minimum width=2cm, minimum height=1cm, fill=black!8] (cpu1) at (0,2) {CPU};
\node[draw, rectangle, minimum width=2cm, minimum height=1cm, fill=black!16] (gpu1) at (3,2) {GPU};
\node[draw, rectangle, minimum width=5cm, minimum height=0.5cm, fill=black!12] (pcie) at (1.5,1) {PCIe 4.0/5.0};
\node at (1.5,2.8) {\footnotesize Discrete (CUDA)};
\draw[<->, thick] (cpu1) -- (gpu1);

% Unified
\node[draw, rectangle, minimum width=2cm, minimum height=1cm, fill=black!8] (cpu2) at (0,-0.5) {CPU};
\node[draw, rectangle, minimum width=2cm, minimum height=1cm, fill=black!16] (gpu2) at (3,-0.5) {GPU};
\node[draw, rectangle, minimum width=5cm, minimum height=0.8cm, fill=black!18] (umem) at (1.5,-1.8) {Unified Memory (400 GB/s)};
\node at (1.5,0.3) {\footnotesize Unified (Apple Silicon)};
\draw[thick] (cpu2.south) -- ++(0,-0.5) -- (umem.north -| cpu2);
\draw[thick] (gpu2.south) -- ++(0,-0.5) -- (umem.north -| gpu2);
\end{tikzpicture}
\caption{Memory architecture comparison. Discrete GPUs require PCIe transfers; unified memory enables zero-copy access.}
\label{fig:memory-model}
\end{figure}
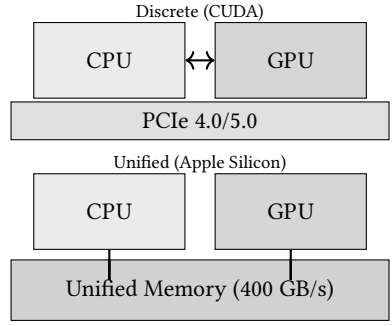

We characterize where unified memory helps and where it does not. In a
discrete-GPU state-vector simulator the state vector is uploaded to device
memory once and then updated in place; individual gates do not trigger
host--device transfers. The PCIe cost is therefore a one-time
upload/read-out of the state (256\,MB for 25 qubits, a few milliseconds on
a PCIe~4.0/5.0 link) plus any host-side inspection of intermediate states.
Unified memory removes this upload/read-out entirely and, more importantly,
enables zero-copy access so the CPU can read or post-process amplitudes
without an explicit copy, and CPU and GPU work can be interleaved at fine
granularity without batching constraints. It does \emph{not} raise peak
arithmetic throughput, nor does it eliminate a per-gate transfer
(production simulators already avoid that). The benefit is largest for
workloads with frequent host--device interaction, such as measurement
sampling, shot-based read-out, and variational outer loops, rather than for
deep, compute-bound circuits. Fig.~\ref{fig:memory-model} illustrates the
architectural difference.

\subsection{Unified-Memory Ablation}

Table~\ref{tab:unified-ablation} reports controlled ablations generated by \path{tools/unified_memory_ablation.py}. The synthetic explicit-copy rows measure two host-memory copies of a complex64 state vector and also report the theoretical PCIe~4.0/5.0 round-trip transfer time for the same byte count. At 25 qubits, the measured host-copy round trip is $34.35\pm20.02$\,ms, while the theoretical PCIe round trip for a 512\,MB transfer is 16.78\,ms at 32\,GB/s and 8.39\,ms at 64\,GB/s. These are small compared with the current 25-qubit circuit runtimes in Table~\ref{tab:pennylane-baseline}; they support the revised claim that unified memory removes setup/read-out copies and programming complexity, not a dominant per-gate transfer.

The host-inspection row compares a 20-qubit six-layer ring-QAOA run with terminal evaluation only against a run that forces a scalar host read after each layer. The measured difference is within run-to-run noise ($1.171\pm0.031$\,s vs. $1.128\pm0.049$\,s), so we do not interpret it as a speedup. The artifact is useful mainly because it bounds this host-read path and shows the methodology for future larger repeat counts.

\begin{table}[t]
\caption{Unified-memory ablation artifacts (mean $\pm$95\% CI)}
\label{tab:unified-ablation}
\centering
\footnotesize
\scriptsize
\setlength{\tabcolsep}{3pt}
\begin{tabular}{@{}p{0.30\linewidth}rp{0.25\linewidth}r@{}}
\toprule
Ablation & Qubits & State size & Mean time \\
\midrule
Explicit round-trip copy & 20 & 8\,MB state / 16\,MB round trip & $1.03\pm0.53$\,ms \\
Explicit round-trip copy & 25 & 256\,MB state / 512\,MB round trip & $34.35\pm20.02$\,ms \\
QAOA terminal evaluation & 20 & 8\,MB state & $1.171\pm0.031$\,s \\
QAOA host read after each layer & 20 & 8\,MB state & $1.128\pm0.049$\,s \\
\bottomrule
\end{tabular}
\end{table}

\section{Hand-Tuned Metal Shaders}\label{sec:shaders}

The pure-MLX tier establishes how far an array framework carries a simulator with no GPU programming at all. This section asks the complementary question: what does hand-tuning buy on the same hardware, the same circuits, and the same convention? Every kernel below lives in one package (\path{src/mlxq/shaders/}), is enabled by a single opt-in flag (\texttt{MLXQ\_METAL\_KERNELS=1}), and is reached through runtime fusion detectors that preserve circuit semantics exactly; with the flag unset the simulator behaves byte-for-byte as the pure tier measured elsewhere in this paper.

\subsection{Kernel Taxonomy}

The kernels exploit one observation in five forms: a structured gate layer that costs $k$ full-state passes on a per-gate path often admits a closed form that costs one.

\textbf{Phase-LUT diagonal layers.} A run of ZZ Trotter terms multiplies each amplitude by a phase that depends only on the number of mismatched neighboring bits, so one pass computes an XOR-popcount parity and indexes a lookup table of $|B|{+}1$ precomputed phases for $|B|$ bonds; no trigonometry and no phase vector in memory. The same machinery covers CZ/CPHASE layers (count of bonds with both bits set, chain/ring closed form $\mathrm{popcount}((i \wedge (i \gg 1)) \wedge m)$), and, grouped by equal angle with per-group LUTs built in double precision, per-bond weighted couplings: the all-to-all long-range Ising layer (300 bonds at 25 qubits, 24 distance groups) and the QPE controlled-power ladder (angles up to $0.4 \cdot 2^{23}$\,rad) each run as ONE pass.

\textbf{GF(2) affine permutation gathers.} X, CNOT, and SWAP act on basis indices as affine maps over GF(2): any run of them sends $|x\rangle$ to $|Mx \oplus c\rangle$ with $M$ invertible. The whole block is therefore a single amplitude permutation, executed as one gather with $n$ popcount-parities per thread ($\mathrm{src}$ bit $k = \mathrm{parity}(\mathrm{rows}_k \wedge y)$); the inverse map composes in Python by replaying the run in reverse, since each such gate is an involution. A 24-gate GHZ CNOT chain becomes one pass; X-only blocks reduce further to $\mathrm{src} = y \oplus c$.

\textbf{Fused tensor-product single-qubit layers.} Single-qubit gates on distinct wires commute, so any window of consecutive single-qubit operations collapses per wire into ONE $2\times2$ product (computed in complex128 at fusion time), and the resulting layer applies $U_a \otimes U_b$ to two qubits per pass: $\lceil n/2 \rceil$ launches instead of one per gate, with identity pairs skipped. Exact collapse tests (rtol $= 0$) recognize algebraic cancellation: an $H,H,X$ window per wire, the Grover diffusion prologue, is literally the matrix $X$ and becomes a single bit-flip gather.

\textbf{Radix-4 butterflies.} The QFT stage kernel performs the Hadamard butterfly and that stage's entire controlled-phase ladder in one pass, using the closed-form ladder angle $\varphi = \pi \cdot \mathrm{low}/2^{s}$; fusing two stages per launch (radix-4, one read and one write per amplitude for both stages) yields the full 25-qubit transform in 13 launches. The same idea gives a radix-4 Walsh--Hadamard kernel: H on four qubits per pass with sign-only arithmetic, $(1/4)\sum_c (-1)^{\mathrm{popcount}(r \wedge c)} a_c$.

\textbf{Basis-conjugated Pauli-pair layers.} All XX bond terms mutually commute, so a uniform XX Trotter layer diagonalizes globally: $H^{\otimes n} \cdot \mathrm{ZZ} \cdot H^{\otimes n}$, with the H layers run by the Walsh kernel. YY conjugates by $V = S H$ (where $V Z V^\dagger = Y$), decomposed into one popcount-LUT diagonal pass for $S^{\otimes n}$ around the same Walsh layers. Cross-family runs are never merged: XX and YY terms on overlapping bonds anticommute, so reordering would change the Trotterized operator.

\subsection{Semantics-Preserving Fusion}

A detector pass over the gate stream routes structure to these kernels without changing circuit semantics. Runs of consecutive commuting diagonal gates (ZZ, CZ, CPHASE; angles free to differ) fuse into LUT or grouped-weighted passes. Runs of X/CNOT/SWAP compose into one GF(2) gather. Windows of single-qubit gates collapse via per-wire matrix products, with identity windows emitting nothing. Same-family XXPHASE/YYPHASE runs conjugate through the ZZ kernel. Wire validation guards every detector (duplicate or out-of-range wires disable fusion), per-qubit-varying parameters fuse through the per-wire product path rather than being assumed uniform, and numerical collapse tests use rtol $= 0$ in complex128 so that a real $R_Z(10^{-6})$ is never silently dropped. Each kernel carries a parity regression test against the pure path; the full suite (253 tests) passes with the flag on and off.

\subsection{Kernel-Level Measurements}

\begin{table}[t]
\caption{Pure-MLX array operations vs.\ hand-written Metal kernels at 25 qubits (mean over ten repeats after one warmup; idle machine; parity to the pure path $\le 5\times10^{-6}$ per test suite)}
\label{tab:shader}
\centering
\footnotesize
\begin{tabular}{@{}lrrrr@{}}
\toprule
Layer & Per-gate & MLX fused & Metal kernel & Metal vs.\ MLX \\
\midrule
ZZ Trotter layer (24 bonds) & \NumZzPerGateMs\,ms & \NumZzFusedMs\,ms & \NumZzMetalMs\,ms & $\NumZzMetalSpeedup\times$ \\
Full QFT (radix-4, 13 launches) & \NumQftPerGateMs\,ms & \NumQftFusedMs\,ms & \NumQftMetalMs\,ms & $\NumQftMetalSpeedup\times$ \\
RX layer (all 25 qubits) & \NumRxPerGateMs\,ms & -- & \NumRxMetalMs\,ms & $\NumRxMetalSpeedup\times$ \\
H layer (fused, all 25 qubits) & \NumHLayerPureMs\,ms & -- & \NumHLayerMetalMs\,ms & $\NumHLayerSpeedup\times$ \\
\bottomrule
\multicolumn{5}{@{}l@{}}{\footnotesize MLX fused ZZ uses a cached phase vector (724\,ms one-time build, amortized across Trotter steps).}\\
\multicolumn{5}{@{}l@{}}{\footnotesize Artifacts: \path{metal_shader_comparison_20260704.json}, \path{shader_suite_20260704.json}; H-layer row from \path{h_layer_clean_20260706.json} (the older JSONs' H entry is marked superseded).}\\
\end{tabular}
\end{table}

Table~\ref{tab:shader} isolates the kernels from end-to-end effects. Where MLX already achieves one pass per layer, hand-tuning buys little: the ZZ kernel sits at the memory floor, $1.5\times$ ahead of the cached-phase-vector multiply. Where a kernel merges structure ACROSS passes, the gap is an order of magnitude: the radix-4 QFT is $12\times$ faster than the fused-ladder MLX path and $3.7\times$ faster than the \texttt{mx.fft} primitive itself ($77$\,ms). The fused Hadamard layer is a more modest $\sim$8$\times$ over sequential structured single-qubit dispatch (its per-gate baseline matches the RX layer of the same shape); like RX it collapses a stack of single-qubit updates into a handful of tensor-product passes, but H has no diagonal shortcut, so the pure-path baseline it improves on is already the structured slice update rather than a dense matmul.

Appendix~\ref{app:sweep} reports the complete per-workload comparison for the full benchmark suite under the paired sweep protocol, and Appendix~\ref{app:hlayer} walks through the Hadamard-layer case line by line, from the pure-MLX single-gate update to the radix-4 Walsh kernel, with the bandwidth accounting that explains the gap.

\subsection{Review-Driven Kernel Development}

Each kernel went through an implement--measure--review--retest loop with an independent code-review model (Codex CLI), and the loop caught real bugs at every level of subtlety: an algebraically wrong radix-4 twiddle derivation in the reviewer's own first proposal (retracted and corrected in its round 2, verified independently before adoption); a silent-correctness hole in which float32 window products under NumPy's default relative tolerance would have dropped $R_Z(10^{-6})$-scale gates as identity (fixed with complex128 products and rtol $= 0$, plus a regression test); and a measured $1.25\times10^{-5}$ phase drift in 300-term float32 phase products (fixed by grouping equal-angle bonds into per-group LUTs built in double precision). The archived reviews ship with the artifact (\path{reviews_shaders_v2/}).

\section{Validation}\label{sec:validation}

\subsection{Regression Test Suite}

We validate correctness through 253 Python tests:

\begin{table}[t]
\caption{Validation test coverage by category}
\label{tab:tests}
\centering
\begin{tabular}{lr}
\toprule
Category & Tests \\
\midrule
Core gates, states \& information primitives & 149 \\
Algorithm \& tutorial example circuits & 41 \\
Optimized-vs-dense internal consistency & 20 \\
MPS backend \& parameter suite & 12 \\
QML wrapper \& QPE energy estimation & 7 \\
Benchmark catalog \& protocol & 3 \\
Cross-framework measurement parity & 1 \\
Visualization plots & 1 \\
Metal shader-kernel parity (incl.\ all-distinct-angle stress) & 19 \\
\midrule
\textbf{Total} & \textbf{253} \\
\bottomrule
\end{tabular}
\end{table}

Each test includes analytical verification against pen-and-paper derivations. Example: for Bell-state preparation $(H \otimes I)\mathrm{CNOT}|00\rangle$, the expected state is $(|00\rangle + |11\rangle)/\sqrt{2}$; our numerical amplitudes match this analytical vector within floating-point tolerance in repeated regression runs.

\subsection{Internal Consistency}

We compare optimized MLX kernels against dense matrix multiplication for identical circuits:

\begin{enumerate}
    \item Define identical circuit structures (gate sequence, qubit count, and parameters)
    \item Execute using optimized MLX kernels
    \item Execute the same circuit with dense reference linear algebra
    \item Compare final states via amplitude $L_2$ distance and probability-space MAE
\end{enumerate}

\begin{table}[t]
\caption{Internal consistency: optimized vs. dense execution}
\label{tab:parity}
\centering
\begin{tabular}{lrr}
\toprule
Circuit & Amplitude $L_2$ & Prob. MAE \\
\midrule
QFT (10q) & $2.3 \times 10^{-7}$ & $1.1 \times 10^{-8}$ \\
Random (10q) & $4.7 \times 10^{-7}$ & $2.8 \times 10^{-8}$ \\
Phase Est. (8q) & $3.1 \times 10^{-7}$ & $1.9 \times 10^{-8}$ \\
Hardware-efficient ansatz (2q) & $5.2 \times 10^{-8}$ & $3.4 \times 10^{-9}$ \\
QAOA (4q) & $1.8 \times 10^{-7}$ & $9.2 \times 10^{-9}$ \\
\bottomrule
\end{tabular}
\end{table}

All deviations are within the expected complex64 floating-point range ($< 10^{-6}$). This internal-consistency check is complemented by the independent complex128 checks below.

The weighted diagonal and ZZ shader layers deserve a dedicated worst case. Their kernels fuse an arbitrary graph of per-bond phases by grouping bonds of equal angle and multiplying one unit phase per group; when every bond carries a \emph{distinct} angle the group count equals the bond count and the kernel performs its longest in-kernel float32 phase product, with no grouping to shorten the chain. Because the per-group phases are precomputed on the host in double precision and the Metal accumulation uses fused multiply--add, this worst case still matches both the pure-MLX path and the analytical complex128 result to $\approx 5\times10^{-7}$ even at $300$ distinct-angle bonds --- an order of magnitude under the $5\times10^{-6}$ parity budget --- and both regimes are pinned by dedicated all-distinct-angle regression tests.

\subsection{Independent complex128 validation}

Table~\ref{tab:precision-validation} reports independent complex128 validation artifacts. The 25-qubit QFT check compares the MLX complex64 state against the analytical complex128 uniform state obtained by applying QFT to $\ket{0\ldots0}$. The 20-qubit ring-QAOA check compares against an independent NumPy complex128 reshape/contract gate path with the same fixed six-layer schedule. The CSV artifact is \path{precision_validation_summary.csv}.

Because parity against the pure-MLX path would not catch a bug shared by both tiers, we additionally validate two representative shader families against \emph{independent} complex128 references rather than the pure path. A 13-qubit long-range Ising weighted-ZZ layer with 78 all-to-all couplings (kernel family S6) matches its analytic reference to infidelity $6.8\times10^{-14}$ and probability MAE $3.5\times10^{-11}$, and an 11-qubit QPE-style weighted CPHASE ladder with per-bond angles up to $205$\,rad (family S4) matches to infidelity $2.0\times10^{-15}$. Both are computed in double precision from the diagonal phase structure, independently of the kernel's own float32 lookup tables, so the agreement exercises the shaders against mathematical ground truth (\path{independent_shader_validation_20260706.json}).

We also validate two full circuits \emph{end to end}, imported from OpenQASM and run with the Metal tier enabled, against an independent from-scratch complex128 statevector simulator; this exercises the OpenQASM parser and the fusion detectors together rather than a single layer. A 9-qubit QPE circuit (\path{qpe_n9.qasm}: H, CPHASE, CCX, CZ) reproduces the independent reference exactly, and a 10-qubit Ising Hamiltonian circuit (\path{ising_n10.qasm}: 480 H/RZ/CNOT gates) matches to infidelity $3.1\times10^{-13}$ (\path{independent_end_to_end_20260706.json}).

\begin{table}[t]
\caption{Independent complex128 validation for representative 20--25 qubit workloads}
\label{tab:precision-validation}
\centering
\scriptsize
\setlength{\tabcolsep}{3pt}
\begin{tabular}{@{}lrrrrr@{}}
\toprule
Workload & Qubits & Amp. $L_2$ & Prob. MAE & Norm drift & Infidelity \\
\midrule
QFT/analytical & 25 & $1.01\times10^{-7}$ & $6.04\times10^{-15}$ & $1.01\times10^{-7}$ & $7.33\times10^{-10}$ \\
Ring-QAOA/NumPy & 20 & $8.93\times10^{-7}$ & $1.09\times10^{-12}$ & $5.01\times10^{-7}$ & $6.06\times10^{-13}$ \\
\bottomrule
\end{tabular}
\end{table}

\subsection{Trotter Error}

Table~\ref{tab:trotter-error} reports a small-system TFIM Trotter-error curve generated by \path{tools/trotter_error_validation.py}. The reference is dense complex128 exact diagonalization using \texttt{scipy.linalg.expm}; the approximate path matches the first-order product-formula schedule used by \texttt{simulate\_hamiltonian}. This is a 6-qubit validation curve, not a 25-qubit exact diagonalization.

\begin{table}[t]
\caption{TFIM first-order Trotter error against exact diagonalization ($n=6$, $J=1$, $h=0.5$, $t=1$)}
\label{tab:trotter-error}
\centering
\footnotesize
\begin{tabular}{rrrr}
\toprule
Steps & State $L_2$ & Prob. MAE & Infidelity \\
\midrule
4 & $2.60\times10^{-1}$ & $4.50\times10^{-4}$ & $4.88\times10^{-2}$ \\
8 & $1.31\times10^{-1}$ & $1.10\times10^{-4}$ & $1.18\times10^{-2}$ \\
16 & $6.56\times10^{-2}$ & $2.74\times10^{-5}$ & $2.90\times10^{-3}$ \\
32 & $3.29\times10^{-2}$ & $6.84\times10^{-6}$ & $7.19\times10^{-4}$ \\
64 & $1.65\times10^{-2}$ & $1.71\times10^{-6}$ & $1.79\times10^{-4}$ \\
\bottomrule
\end{tabular}
\end{table}

\subsection{Framework Alignment}

We reproduce benchmark families inspired by PennyLane, Yao.jl, and Qulacs specifications~\cite{PennyLane2022,YaoJL2024,Qulacs}, matching the intended circuit structure, layer counts, and parameter conventions where the local implementation supports them. These runs are workload-alignment artifacts. Same-machine PennyLane \texttt{lightning.qubit} and Qiskit Aer CPU timing baselines are reported in Sec.~\ref{sec:pennylane-baseline}; broader cross-framework correctness and performance baselines remain future work.

\section{Discussion}\label{sec:discussion}

\subsection{Unified Memory Trade-offs}

\textbf{Advantages}:
\begin{enumerate}
    \item Zero-copy access eliminates explicit setup/read-out copies
    \item Simplified programming without explicit memory management
    \item Flexible CPU-GPU scheduling without batching constraints
    \item Full memory pool available to both processors
\end{enumerate}

\textbf{Limitations}:
\begin{enumerate}
    \item Peak GPU throughput on discrete architectures exceeds unified memory bandwidth
    \item Deep, compute-bound circuits may favor discrete GPUs
    \item Maximum capacity (512\,GB) trails specialized HPC nodes
\end{enumerate}

For typical local NISQ-era exploration with moderate depth, unified memory can improve workflow simplicity and reduce setup/read-out copying. The ablation in Table~\ref{tab:unified-ablation} shows that explicit setup/read-out copy costs for a 25-qubit state are in the millisecond-to-tens-of-milliseconds range, not the dominant cost in the measured circuits. The unified-memory claim should therefore be read as a workflow and integration claim rather than a measured throughput advantage. Throughput comparisons remain workload- and hardware-dependent and require co-located baselines.

\subsection{Educational Applications}

\mlxquantum significantly lowers barriers for quantum computing education, enabling hands-on algorithm exploration in classroom settings without specialized infrastructure:

\textbf{Zero Configuration}: Unlike CUDA-based frameworks requiring driver installation, GPU compatibility verification, and environment configuration, \mlxquantum runs immediately on any Apple Silicon Mac with a single \texttt{pip install} command.

\textbf{Familiar Semantics}: The Python API mirrors NumPy conventions, enabling students familiar with scientific Python to begin quantum programming without learning new paradigms. State vectors behave as arrays; gates apply as transformations.

\textbf{Accessible Hardware}: MacBook Air and Pro models, standard in many university computer labs and student ownership, provide sufficient capability for educational quantum simulation up to 20+ qubits.

\textbf{Comprehensive Tutorials}: The framework includes 12 complete tutorials covering:
\begin{itemize}
    \item Quantum Circuit Born Machine (QCBM) for generative modeling
    \item Quantum Fourier Transform with product representation derivation
    \item GHZ state preparation and entanglement verification
    \item Hamiltonian evolution with Trotter error analysis
    \item Hardware-efficient ansatz optimization; molecular Hamiltonian workflows require future fermion-to-qubit mapping support
    \item QAOA for MaxCut optimization
\end{itemize}

Each tutorial provides mathematical foundations, pen-and-paper derivations, quantum circuit diagrams, and executable Python code. The 62 worked examples span fundamental concepts (superposition, entanglement, measurement) through advanced algorithms (Grover search, phase estimation, variational optimization).

\textbf{Comprehensive Documentation}: The \mlxquantum documentation and tutorials provide extensive educational material including validation suite documentation, algorithm tutorials with step-by-step derivations, and appendices covering quantum information theory fundamentals. This reference serves both as a learning resource for students new to quantum computing and as a detailed technical reference for experienced practitioners.

\textbf{Interactive Benchmark Runner}: We provide both a terminal runner and the QuantumStudio desktop UI so researchers and students can run benchmarks without manual shell scripting. The runner supports:
\begin{itemize}
    \item Interactive selection of algorithm benchmarks
    \item Configurable qubit ranges and iteration counts
    \item Configurable warmup and measured-repeat counts for reproducible timing
    \item Progress reporting during benchmark runs
    \item Automatic CSV/JSON export, raw timing distributions, timing summaries, and run manifests for further analysis
    \item Hardware detection and capability reporting
\end{itemize}

This accessibility-first design ensures that quantum algorithm exploration is available to the broadest possible audience, from undergraduate coursework to graduate research.

\subsection{Comparison with Prior Work}

Table~\ref{tab:comparison} positions \mlxquantum within the ecosystem:

\begin{table}[t]
\caption{Framework comparison and baseline status}
\label{tab:comparison}
\centering
\scriptsize
\setlength{\tabcolsep}{2.5pt}
\begin{tabular}{llll}
\toprule
Framework & Native target & User API & Status in this work \\
\midrule
cuQuantum & CUDA GPU & C++/Python & Cited; not co-located \\
qsim & CPU/CUDA & C++/Python & Cited; not co-located \\
QuEST & CPU/GPU/MPI & C & Cited; not co-located \\
PennyLane & Plugin backends & Python & \texttt{lightning.qubit} measured \\
Qiskit Aer & CPU/CUDA & Python & CPU statevector measured \\
\textbf{Qupertino} & \textbf{MLX/Metal} & \textbf{Python} & \textbf{Measured artifact} \\
\bottomrule
\end{tabular}
\end{table}

\mlxquantum targets MLX/Metal on Apple Silicon and provides a CUDA-free local workflow for that hardware family. Table~\ref{tab:comparison} should not be read as a general throughput ranking: the measured relation in Table~\ref{tab:pennylane-baseline} is workload- and size-dependent, with \mlxquantum fastest at 20--25 qubits on most gate-identical workloads and the CPU baselines fastest on shallow circuits at small sizes.

Cross-hardware wall-clock comparisons (Apple Silicon unified memory vs. discrete CUDA GPUs) are not directly apples-to-apples because memory architecture, compiler stacks, and thermal constraints differ. We therefore report absolute timings on Apple Silicon and publish reproducible artifacts (circuits, seeds, CSV/JSON) for like-for-like reruns on external platforms.

\subsection{QASMBench Circuit Integration}

Table~\ref{tab:qasmbench} shows representative QASMBench circuits executed by \mlxquantum, demonstrating OpenQASM 2.0 compatibility across circuit scales.

\begin{table}[t]
\caption{QASMBench circuit execution on M1 Max (single cold execution, incl. first-call kernel setup)}
\label{tab:qasmbench}
\centering
\footnotesize
\begin{tabular}{lrrr}
\toprule
Circuit & Qubits & Gates & Time (ms) \\
\midrule
bell.qasm & 2 & 2 & 1.65 \\
grover\_2qubit.qasm & 2 & 12 & 1.31 \\
qft\_n4.qasm & 4 & 12 & 0.83 \\
teleportation\_n3.qasm & 3 & 8 & 0.71 \\
deutsch\_n2.qasm & 2 & 5 & 0.65 \\
inverseqft\_n4.qasm & 4 & 8 & 0.61 \\
qec9xz\_n17.qasm & 17 & 53 & 3.75 \\
multiplier\_n15.qasm & 15 & 70 & 5.59 \\
dnn\_n16.qasm & 16 & 2,016 & 163.4 \\
qft\_n18.qasm & 18 & 783 & 37.8 \\
\bottomrule
\end{tabular}
\end{table}

\subsection{Performance Analysis Model}

It is important to state the cost model accurately, because a naive
per-gate transfer model overstates the benefit. For a circuit with $G$
gates on an $n$-qubit state, a competently implemented discrete-GPU
simulator uploads the state once, applies all $G$ gates in device memory,
and reads the result back:
\begin{equation}
T_{\text{discrete}} = t_{\text{transfer}}^{\text{setup}} + \sum_{i=1}^{G} t_{\text{compute},i} + t_{\text{transfer}}^{\text{readout}},
\end{equation}
where $t_{\text{transfer}}^{\text{setup}} + t_{\text{transfer}}^{\text{readout}} = 2 \cdot 8 \cdot 2^n / B_{\text{PCIe}}$
for a complex64 state. On a contemporary PCIe~4.0/5.0 link
($B_{\text{PCIe}} \approx 32$--$64$\,GB/s) this is a few milliseconds for a
25-qubit (256\,MB) state, and it is a one-time cost rather than a per-gate one. The
unified-memory model removes these two transfer terms:
\begin{equation}
T_{\text{unified}} = \sum_{i=1}^{G} t'_{\text{compute},i}.
\end{equation}

For deep circuits the transfer term is thus negligible and both models are
dominated by compute; unified memory offers no asymptotic runtime advantage
there. The transfer term becomes significant only when the host must touch
the state frequently, as in per-shot measurement, mid-circuit read-out, and
variational outer loops that move amplitudes or expectation values back to
the CPU each iteration. In those settings, avoiding repeated copies,
together with the elimination of explicit memory management, is the
structural benefit unified memory provides. The ablation in Table~\ref{tab:unified-ablation} provides
the first bounded measurement of explicit copies and forced host-side
amplitude reads; larger repeat counts and additional host-interaction-heavy
workloads remain useful future measurements.

A first-order traffic model indicates how close the dispatch path runs to
the memory wall. The 25-qubit complex64 state occupies $2^{25} \times 8$\,B
$= 268$\,MB ($0.268$\,GB), and every structured kernel reads and writes the
full state at least once, giving a per-gate traffic floor of $2 \times
0.268 = 0.537$\,GB. The 325-gate QFT therefore moves at least $325 \times
0.537 \approx 174$\,GB; completing in $0.721$\,s, it sustains roughly
240\,GB/s, or about $60\%$ of the M1 Max's 400\,GB/s peak bandwidth. The measured runtime therefore sits within a small factor of the
memory-bandwidth bound, which is consistent with the elementwise,
transpose-free structure of the dispatched kernels and leaves limited
headroom for further single-gate optimization; the remaining gains would
have to come from gate fusion, which reduces the number of full-state
passes rather than the cost of each pass.

Section~\ref{sec:shaders} tests that prediction directly. A runtime fusion pass plus a full tier of hand-written Metal kernels reduce the number of full-state passes per structured layer, and the measured gains land exactly where the model says they must: order-of-magnitude end-to-end speedups where a kernel merges structure across passes, and no gain where the pure-MLX operation already sits at the memory floor.

\subsection{Failure Modes and Capacity Outlook}

Higher-memory Apple Silicon systems increase state-vector capacity headroom, but this manuscript does not report measured M2 Ultra or M3 Ultra wall-clock results, and earlier projection tables have been removed for that reason. The only quantitative scaling claims in the main text are therefore the M1 Max measurements and the memory-footprint arithmetic in Table~\ref{tab:memory-capacity}, and larger-memory timing should be treated as a future validation target rather than an extrapolation from the present data.

Table~\ref{tab:failure-modes} summarizes concrete failure modes and scope boundaries generated by \path{tools/failure_modes_report.py}. The capacity model assumes complex64 state vectors on a 32\,GB M1 Max: a single state first exceeds 32\,GB at 32 qubits, while a conservative 4$\times$ workspace model first exceeds 32\,GB at 30 qubits. This explains why the measured state-vector evidence is centered on 20--25 qubits and why 30+ qubit entries are not reported as validated performance.

\begin{table}[t]
\caption{Failure Modes and scope boundaries}
\label{tab:failure-modes}
\centering
\footnotesize
\scriptsize
\setlength{\tabcolsep}{3pt}
\begin{tabular}{@{}p{0.23\linewidth}p{0.18\linewidth}p{0.53\linewidth}@{}}
\toprule
Mode & Status & Finding \\
\midrule
Out-of-memory / swap & Capacity-bounded & 4$\times$ workspace exceeds 32\,GB at 30q; single state exceeds 32\,GB at 32q \\
Small-qubit overhead & Known risk & Python dispatch and MLX graph/synchronization overhead can dominate small circuits \\
OpenQASM semantics & Diagnosed & Unitary subset executes; measure/reset/if/barrier/opaque semantics are outside scope \\
Precision drift & Measured subset & QFT 25q and ring-QAOA 20q complex128 checks reported in Table~\ref{tab:precision-validation} \\
Variational convergence & Scope-limited & Ansatz optimization is not a molecular VQE convergence claim \\
\bottomrule
\end{tabular}
\end{table}

\subsection{Threats to Validity}\label{sec:threats}
We organize the risks to our conclusions in the three standard categories for empirical systems work.

\textbf{Internal validity} (does the measurement support the causal claim?). Timing is confounded by session-level thermal and memory-pressure drift, which we control by interleaving all backends round-robin within every repeat, by two warmups, and by a quietness probe; we report paired per-repeat ratios rather than ratio-of-means. Two residual risks remain: backend order within a repeat is fixed rather than randomized, and a minority of kernel-level cells are high-variance (the all-H micro-benchmark of App.~\ref{app:hlayer} is the clearest, which is why we quote a re-measured idle-machine figure there). The dense-path ablation isolates kernel specialization from GPU execution as a controlled within-session comparison.

\textbf{External validity} (do the results generalize?). All measurements are from one M1 Max host, complex64 state vectors, a fixed set of gate-identical workloads, and pinned simulator versions (Qiskit 2.4.2, qiskit-aer 0.17.2, PennyLane 0.43.0, MLX 0.30.6). Results may not transfer to other Apple-Silicon generations, to circuit families outside the tested set, or to other simulator releases; we give M4 Max bandwidth as context but do not measure it. No discrete-GPU or CUDA/cuStateVec baseline is co-located.

\textbf{Construct validity} (are the metrics the right proxies?). Wall-clock time is the practitioner-relevant metric for a local simulator, and we report it with intervals rather than single runs. Shader correctness is established primarily by parity against the pure-MLX path, which a bug shared by both paths would not catch; independent complex128 anchors for four gate families (QFT, ring-QAOA, a long-range Ising weighted-ZZ layer, and a QPE-style weighted-CPHASE ladder) together with the exact-diagonalization Trotter curve mitigate this, and extending the anchors to OpenQASM-imported circuits and the remaining shader families is future work (Sec.~\ref{sec:limitations}).

\subsection{Limitations and Future Work}\label{sec:limitations}

\textbf{Scale}: State-vector simulation is limited to $\sim$30--35 qubits by memory capacity. A preliminary MPS backend extends measured coverage to 100--150 qubits for limited-entanglement workloads (Sec.~\ref{sec:mps-scaleout}) as a scale-feasibility probe; its bond-dimension and truncation-error instrumentation is still being hardened (the GHZ summary currently records \texttt{bond\_max}=0 rather than the analytic value 2), so we treat these figures as indicative reach rather than a validated contribution. Exact simulation of highly entangled circuits beyond the state-vector ceiling remains out of scope.

\textbf{Performance}: We do not claim computational parity with highly optimized CUDA implementations on discrete GPUs; no co-located CUDA baseline is reported. On the same M1 Max host, the Metal shader tier is fastest by mean runtime in all 18 four-way comparison cells, and the GPU-backed pure-MLX tier is fastest among the same-host CPU baselines (Aer, \texttt{lightning.qubit}) on all six workloads at 25 qubits by mean --- five of six by the stricter paired-ratio test, with the Grover proxy a statistical tie against Aer; small gate-sparse circuits at 15 qubits favor the CPU baselines. Because the pure tier still executes on the GPU, ``same-host CPU baseline'' denotes the comparison set, not a claim of CPU-only execution. per-kernel launch overhead still dominates shallow circuits at small qubit counts, wall-clock varies with unified-memory pressure between sessions (which is why all comparative claims are within-session paired ratios), and \texttt{mx.fft} kernels are unavailable for some register sizes ($2^{21}$, $2^{22}$ in MLX 0.30.6; the QFT primitive falls back to the fused circuit ladder there). The density-matrix steady-state path and generic dense multi-qubit gates retain further optimization headroom.

\textbf{Baseline coverage}: This revision reports PennyLane \texttt{lightning.qubit}, PennyLane \texttt{lightning.kokkos} (OpenMP, Table~\ref{tab:kokkos}), and Qiskit Aer CPU baselines on the same M1 Max host. CUDA/cuStateVec baselines still require a separate CUDA host and are not reported as co-located Apple-Silicon comparisons; no Metal-accelerated PennyLane or Qiskit backend exists.

\textbf{Statistical protocol}: Interleaving controls session drift, but the backend order within each repeat is fixed rather than randomized, so residual order effects are not averaged out; randomized-order reruns are future work. A minority of kernel-level and metal-tier cells are high-variance --- the all-H micro-benchmark (Sec.~\ref{app:hlayer}) and a few shader-sweep rows show first-launch GPU-graph-compilation outliers (for example an isolated $399.9$\,ms phase-estimation metal run against a $\sim$60\,ms steady state) --- and the 29-workload sweep in App.~\ref{app:sweep} reports paired ranges rather than confidence intervals. The large Metal-versus-CPU margins are robust to this, but the pure-tier and Kokkos margins are the ones that warrant the cautious, interval-qualified wording used above; where a paired interval spans parity we do not count the cell as a win.

\textbf{Correctness breadth}: Shader correctness is established primarily by parity against the pure-MLX path, which would not catch a bug shared by both paths. The independent complex128 anchors now cover four gate families --- QFT (25q), ring-QAOA (20q), a long-range Ising weighted-ZZ layer (13q, 78 couplings), and a QPE-style weighted-CPHASE ladder (11q) --- alongside the exact-diagonalization Trotter curve; extending independent references to OpenQASM-imported circuits and the remaining shader families is the natural next validation step.

\textbf{Numerical precision}: The implementation currently uses complex64 state vectors. This revision adds independent complex128 checks for QFT at 25 qubits and ring-QAOA at 20 qubits, plus a 6-qubit TFIM Trotter-error curve. Future validation should expand the same metrics to QCBM and larger Hamiltonian workloads where exact references remain feasible.

\textbf{Unified-memory ablation}: This revision includes synthetic explicit-copy and forced host-read ablations. These do not establish a throughput advantage; they bound setup/read-out and host-inspection costs for the tested cases.

\textbf{Variational convergence}: Hardware-efficient ansatz timings are optimization-cost diagnostics. They do not demonstrate molecular ground-state accuracy, fermion-to-qubit mappings, or convergence against exact diagonalization.

\textbf{Failure modes}: The current failure-mode report covers memory thresholds, small-qubit overhead risk, OpenQASM semantic scope, precision-drift checks, and variational-convergence scope. It does not yet include live swap-thrash stress tests or stochastic convergence sweeps across many seeds.

\textbf{Energy Efficiency}: Power/energy measurements are not reported in this work. Future study using macOS \texttt{powermetrics} instrumentation with matched workloads will characterize sustainability metrics for quantum simulation workloads on consumer hardware.

\textbf{Tensor Networks}: The MPS backend is measured here to 100--150 qubits at bond cap 64 (Sec.~\ref{sec:mps-scaleout}); raising the bond ceiling toward 256, adding two-dimensional tensor-network geometries, and quantifying truncation error against exact references at small sizes are the natural next steps.

\textbf{Noise Models}: Integration of depolarizing, amplitude damping, and phase damping channels for NISQ-era noise simulation is planned.

\section{Related Work}\label{sec:related}

NVIDIA's cuQuantum~\cite{cuQuantum2023} provides GPU-optimized libraries for quantum state-vector and tensor network simulation, achieving high throughput on NVIDIA GPUs through CUDA kernels and multi-GPU scaling with cuStateVec and cuTensorNet APIs. Google's qsim~\cite{Qsim} implements aggressive gate fusion and AVX/FMA vectorization for CPU backends alongside GPU acceleration, focusing on simulation efficiency for random circuit sampling benchmarks. QuEST~\cite{QuEST,Jones2019quest} offers hybrid CPU-GPU deployment with automatic hardware selection, supporting both state-vector and density matrix simulations across diverse computing environments. Intel-QS~\cite{IntelQS} targets distributed computing with MPI support for multi-node simulation, avoiding explicit gate matrix representations for memory efficiency. At the extreme end of scale, supercomputer campaigns have simulated 45-qubit circuits using half a petabyte of distributed memory~\cite{HanerSteiger2017}, and long-running massively parallel simulators document how memory, communication, and precision trade off at that scale~\cite{DeRaedt2019}; these efforts bracket the opposite end of the design space from the single-node consumer hardware targeted here. Against this ecosystem, our work is complementary: it contributes a Python-first MLX/Metal artifact with OpenQASM import, local run manifests, and QuantumStudio UI integration rather than a controlled memory-hierarchy study or a CUDA-throughput competitor.

On the programming-framework side, PennyLane~\cite{PennyLane2022} provides differentiable hybrid quantum-classical computation with automatic differentiation through multiple backend devices, enabling gradient-based optimization for variational algorithms. Qiskit~\cite{Qiskit2023} offers comprehensive tooling integrated with IBM quantum hardware, including Qiskit Aer for high-performance simulation. Cirq~\cite{Cirq2024} targets NISQ algorithm prototyping with Google hardware integration. Yao.jl~\cite{YaoJL2024} delivers extensible tensor-network simulation in Julia with block-structured circuit abstractions.

Benchmarking standards shape how results such as ours can be compared. QASMBench~\cite{QASMBench2020} provides standardized OpenQASM circuits spanning quantum chemistry, cryptography, and optimization domains with small (2--10 qubits), medium (11--27 qubits), and large (28--127 qubits) scale categories. SupermarQ~\cite{SupermarQ2022} and MQTBench~\cite{MQTBench2022} offer scalable quantum benchmark suites with configurable circuit parameters. Benchpress~\cite{Benchpress2024} from IBM Quantum provides comprehensive performance characterization methodology aligning with community standards.

\mlxquantum complements these frameworks by providing CUDA-free deployment on consumer Apple Silicon hardware with unified memory optimization, filling a gap in the quantum system software ecosystem for researchers and educators in the Apple hardware ecosystem.

\section{Conclusion}\label{sec:conclusion}

We have presented a Python-first quantum simulation framework for Apple Silicon that measures, on one machine and one set of gate-identical circuits, both how far an array framework carries a state-vector simulator and what hand-tuned Metal shaders add on top. The pure-MLX tier, structured-gate dispatch expressed entirely in array operations, executes moderate-scale workloads competitively without any GPU programming: in the four-way interleaved campaign the GPU-backed pure tier is fastest among the same-host CPU baselines on all six 25-qubit workloads by mean runtime (five of six by the stricter paired-ratio test, with Grover tied against Aer). The shader tier answers the second question quantitatively: seven kernel families plus semantics-preserving fusion detectors make it fastest by mean in all 18 comparison cells, with 25-qubit paired ratios of $23$--$47\times$ over Qiskit Aer CPU and $19$--$95\times$ over PennyLane \texttt{lightning.qubit}; gate-stream QFT completes in $59$\,ms, and an isolated radix-4 stage kernel also beats MLX's own \texttt{mx.fft} primitive under matched bit-reversed, no-final-swap semantics, while the full-suite paired sweep shows up to $25\times$ over the pure tier with the gains landing exactly where kernels merge structure across full-state passes. The contribution is an MLX-native two-tier simulator, a review-driven kernel development record, and a reproducibility workflow in which the headline figures are single-sourced from machine-readable artifacts and regenerated, behind a drift gate, on every build.

Our validation suite of 253 regression tests with analytical verification provides useful internal regression coverage across gate operations, quantum information primitives, variational circuits, and Hamiltonian simulation methods. OpenQASM 2.0 import capabilities support reproducible execution of QASMBench circuits, while workload alignment with PennyLane, Yao.jl, and Qulacs specifications~\cite{PennyLane2022,YaoJL2024,Qulacs} facilitates future like-for-like comparison. This revision adds independent complex128 checks for QFT at 25 qubits and ring-QAOA at 20 qubits, plus a small-system TFIM Trotter-error curve against exact diagonalization.

The unified memory architecture provides practical workflow advantages for quantum simulation: zero-copy access avoids explicit setup/read-out copies, simplified programming removes manual memory management, and shared arrays allow CPU--GPU interoperation during interactive exploration. These benefits are most relevant for workloads with frequent host-side interaction, such as measurement sampling and variational outer loops. They do not imply computational parity with optimized discrete-GPU simulators, and they do not change the asymptotic qubit limit of exact state-vector simulation; memory requirements remain exponential ($O(2^n)$).

The measured comparison is deliberately scoped: six gate-identical workloads on one M1 Max host against CPU backends, with shallow small-qubit circuits still favoring the baselines and no co-located CUDA comparison. Within that scope, \mlxquantum is the fastest same-machine option at 20--25 qubits on most workloads, and it remains an MLX-native quantum-simulation artifact with reproducible benchmark outputs, a CUDA-free local workflow, OpenQASM import, QuantumStudio UI integration, and educational examples for Apple Silicon users.

\textbf{Future Directions}: Planned extensions include CUDA/cuStateVec baselines on a comparable discrete-GPU host, broader independent complex128 checks for additional workloads, larger host-inspection ablation sweeps, Matrix Product State tensor network methods for larger approximate simulation, quantum error correction circuit simulation for fault-tolerant algorithm validation, stochastic convergence sweeps, and QASMBench deep-circuit support. Larger-memory Apple Silicon systems should be treated as future validation targets, not as evidence for the current measured claims.

The \mlxquantum framework is open-source and available for community contribution. Its current role is a reproducible Apple Silicon software-systems artifact for quantum algorithm development, education, and local benchmarking.

\begin{acks}
Portions of the software engineering, benchmarking harness, and manuscript preparation were carried out with the assistance of AI-based coding and writing tools; the author reviewed, validated, and takes full responsibility for all code, measurements, and text, in accordance with the ACM Policy on Authorship.
\end{acks}

\section*{Competing Interests}
The author declares no competing interests.

\section*{Data Availability}

Source code, benchmark scripts, QuantumStudio UI tooling, and CSV/JSON performance artifacts are available at \url{https://github.com/BoltzmannEntropy/Qupertino}. Project website: \url{https://boltzmannentropy.github.io/QupertinoWEB/}. The revision evidence artifacts are generated under \path{paper/tqc-acm-2026/evidence_artifacts/}. The files used directly in this manuscript are:
\begin{quote}\footnotesize
\path{interleaved_4way_20260704/} (four-way comparison campaign; Tables~\ref{tab:pennylane-baseline} and paired ratios)\\
\path{kokkos_campaign_20260704/} (Table~\ref{tab:kokkos})\\
\path{shader_sweep_20260704/} (full 29-workload shader sweep; App.~\ref{app:sweep})\\
\path{qasm_shader_sweep_20260704/} (OpenQASM import sweep)\\
\path{metal_shader_comparison_20260704.json}, \path{shader_suite_20260704.json}, \path{h_layer_clean_20260706.json} (Table~\ref{tab:shader})\\
\path{precision_validation_summary.csv}, \path{independent_shader_validation_20260706.json}, \path{independent_end_to_end_20260706.json}, \path{trotter_error_summary.csv}\\
\path{aer_sensitivity_20260706.json} (Aer optimization-level sensitivity)\\
\path{unified_memory_ablation_summary.csv}, \path{qiskit_aer_baseline_summary.csv}\\
\path{failure_modes.json}, \path{claim_audit.json}, \path{claim_audit.md}
\end{quote}
The manuscript's headline figures --- test counts, the paired backend/Metal ratios, the kernel-cell timings of Table~\ref{tab:shader}, sweep counts, and the dispatch-ablation ratios --- are single-sourced from these artifacts: \path{tools/gen_paper_numbers.py} derives each figure once and emits \path{generated_numbers.tex} (one macro per figure, \texttt{\textbackslash input} into the preamble), so every single-sourced figure is regenerated from the data rather than typed by hand. \path{build.sh} runs this regeneration and a drift check before every compile, and \path{claim_audit.json}/\path{.md}, regenerated against this manuscript, record each figure's value and source artifact; broadening the automated audit to the remaining hardcoded prose numbers is in progress. Each central campaign manifest (four-way, Kokkos, shader-sweep, QASM-sweep) carries a provenance block with the git commit, package versions, command line, environment flags, and synchronization sequence.
Results are reproducible on Apple Silicon hardware with Python 3.11.3, MLX 0.30.6, and recorded run manifests. A DOI-backed archival deposit remains a camera-ready packaging task.

\bibliographystyle{ACM-Reference-Format}
\bibliography{references_corrected}


\begin{thebibliography}{33}

%%% the Web address.

\ifx \showCODEN    \undefined \def \showCODEN     #1{\unskip}     \fi
\ifx \showDOI      \undefined \def \showDOI       #1{#1}\fi
\ifx \showISBNx    \undefined \def \showISBNx     #1{\unskip}     \fi
\ifx \showISBNxiii \undefined \def \showISBNxiii  #1{\unskip}     \fi
\ifx \showISSN     \undefined \def \showISSN      #1{\unskip}     \fi
\ifx \showLCCN     \undefined \def \showLCCN      #1{\unskip}     \fi
\ifx \shownote     \undefined \def \shownote      #1{#1}          \fi
\ifx \showarticletitle \undefined \def \showarticletitle #1{#1}   \fi
\ifx \showURL      \undefined \def \showURL       {\relax}        \fi
% invisible to TeX
\providecommand\bibfield[2]{#2}
\providecommand\bibinfo[2]{#2}
\providecommand\natexlab[1]{#1}
\providecommand\showeprint[2][]{arXiv:#2}

\bibitem[Aleksandrowicz et~al\mbox{.}(2023)]%
        {Qiskit2023}
\bibfield{author}{\bibinfo{person}{Gadi Aleksandrowicz},
  \bibinfo{person}{Thomas Alexander}, \bibinfo{person}{Panagiotis Barkoutsos},
  \bibinfo{person}{Luciano Bello}, \bibinfo{person}{Yael Ben-Haim},
  \bibinfo{person}{David Bucher}, {et~al\mbox{.}}}
  \bibinfo{year}{2023}\natexlab{}.
\newblock \bibinfo{title}{Qiskit: An Open-source Framework for Quantum
  Computing}.
\newblock
\newblock
\urldef\tempurl%
\url{https://doi.org/10.5281/zenodo.2562111}
\showDOI{\tempurl}


\bibitem[{Apple Inc.}(2024)]%
        {Apple2024M4}
\bibfield{author}{\bibinfo{person}{{Apple Inc.}}}
  \bibinfo{year}{2024}\natexlab{}.
\newblock \bibinfo{title}{Apple introduces {M4 Pro} and {M4 Max}}.
\newblock
\newblock
\urldef\tempurl%
\url{https://www.apple.com/newsroom/2024/10/apple-introduces-m4-pro-and-m4-max/}
\showURL{%
\tempurl}


\bibitem[{Apple Machine Learning Research}(2024)]%
        {MLXFramework2024}
\bibfield{author}{\bibinfo{person}{{Apple Machine Learning Research}}.}
  \bibinfo{year}{2024}\natexlab{}.
\newblock \bibinfo{title}{{MLX}: An Array Framework for {Apple Silicon}}.
\newblock
\newblock
\urldef\tempurl%
\url{https://ml-explore.github.io/mlx/build/html/index.html}
\showURL{%
\tempurl}


\bibitem[Bayraktar et~al\mbox{.}(2023)]%
        {cuQuantum2023}
\bibfield{author}{\bibinfo{person}{Harun Bayraktar}, \bibinfo{person}{Ali
  Charara}, \bibinfo{person}{David Clark}, \bibinfo{person}{Saul Cohen},
  \bibinfo{person}{Timothy Costa}, \bibinfo{person}{Yao-Lung~L. Fang},
  \bibinfo{person}{Yang Gao}, \bibinfo{person}{Jack Guan},
  \bibinfo{person}{John Gunnels}, \bibinfo{person}{Azzam Haidar},
  {et~al\mbox{.}}} \bibinfo{year}{2023}\natexlab{}.
\newblock \showarticletitle{cuQuantum {SDK}: A High-Performance Library for
  Accelerating Quantum Science}.
\newblock \bibinfo{journal}{\emph{arXiv preprint arXiv:2308.01999}}
  (\bibinfo{year}{2023}).
\newblock


\bibitem[Bergholm et~al\mbox{.}(2022)]%
        {PennyLane2022}
\bibfield{author}{\bibinfo{person}{Ville Bergholm}, \bibinfo{person}{Josh
  Izaac}, \bibinfo{person}{Maria Schuld}, \bibinfo{person}{Christian Gogolin},
  \bibinfo{person}{Shahnawaz Ahmed}, \bibinfo{person}{Vishnu Ajith},
  \bibinfo{person}{M~Sohaib Alam}, \bibinfo{person}{Guillermo Alonso-Linaje},
  \bibinfo{person}{B AkashNarayanan}, \bibinfo{person}{Ali Asadi},
  {et~al\mbox{.}}} \bibinfo{year}{2022}\natexlab{}.
\newblock \showarticletitle{{PennyLane}: Automatic differentiation of hybrid
  quantum-classical computations}.
\newblock \bibinfo{journal}{\emph{arXiv preprint arXiv:1811.04968}}
  (\bibinfo{year}{2022}).
\newblock


\bibitem[Childs et~al\mbox{.}(2021)]%
        {Childs2019}
\bibfield{author}{\bibinfo{person}{Andrew~M. Childs}, \bibinfo{person}{Yuan
  Su}, \bibinfo{person}{Minh~C. Tran}, \bibinfo{person}{Nathan Wiebe}, {and}
  \bibinfo{person}{Shuchen Zhu}.} \bibinfo{year}{2021}\natexlab{}.
\newblock \showarticletitle{Theory of {Trotter} Error with Commutator Scaling}.
\newblock \bibinfo{journal}{\emph{Physical Review X}}  \bibinfo{volume}{11}
  (\bibinfo{year}{2021}), \bibinfo{pages}{011020}.
\newblock
\urldef\tempurl%
\url{https://doi.org/10.1103/PhysRevX.11.011020}
\showDOI{\tempurl}


\bibitem[Coppersmith(1994)]%
        {Coppersmith1994}
\bibfield{author}{\bibinfo{person}{Don Coppersmith}.}
  \bibinfo{year}{1994}\natexlab{}.
\newblock \bibinfo{booktitle}{\emph{An approximate {F}ourier transform useful
  in quantum factoring}}.
\newblock \bibinfo{type}{{T}echnical {R}eport} RC~19642.
  \bibinfo{institution}{IBM Research}.
\newblock
\newblock
\shownote{arXiv:quant-ph/0201067}.


\bibitem[Cross et~al\mbox{.}(2017)]%
        {OpenQASM2017}
\bibfield{author}{\bibinfo{person}{Andrew~W. Cross}, \bibinfo{person}{Lev~S.
  Bishop}, \bibinfo{person}{John~A. Smolin}, {and} \bibinfo{person}{Jay~M.
  Gambetta}.} \bibinfo{year}{2017}\natexlab{}.
\newblock \showarticletitle{Open Quantum Assembly Language}.
\newblock \bibinfo{journal}{\emph{arXiv preprint arXiv:1707.03429}}
  (\bibinfo{year}{2017}).
\newblock


\bibitem[De~Raedt et~al\mbox{.}(2019)]%
        {DeRaedt2019}
\bibfield{author}{\bibinfo{person}{Hans De~Raedt}, \bibinfo{person}{Fengping
  Jin}, \bibinfo{person}{Dennis Willsch}, \bibinfo{person}{Madita Willsch},
  \bibinfo{person}{Naoki Yoshioka}, \bibinfo{person}{Nobuyasu Ito},
  \bibinfo{person}{Shengjun Yuan}, {and} \bibinfo{person}{Kristel Michielsen}.}
  \bibinfo{year}{2019}\natexlab{}.
\newblock \showarticletitle{Massively parallel quantum computer simulator,
  eleven years later}.
\newblock \bibinfo{journal}{\emph{Computer Physics Communications}}
  \bibinfo{volume}{237} (\bibinfo{year}{2019}), \bibinfo{pages}{47--61}.
\newblock
\urldef\tempurl%
\url{https://doi.org/10.1016/j.cpc.2018.11.005}
\showDOI{\tempurl}


\bibitem[Farhi et~al\mbox{.}(2014)]%
        {QAOA2014}
\bibfield{author}{\bibinfo{person}{Edward Farhi}, \bibinfo{person}{Jeffrey
  Goldstone}, {and} \bibinfo{person}{Sam Gutmann}.}
  \bibinfo{year}{2014}\natexlab{}.
\newblock \showarticletitle{A quantum approximate optimization algorithm}.
\newblock \bibinfo{journal}{\emph{arXiv preprint arXiv:1411.4028}}
  (\bibinfo{year}{2014}).
\newblock


\bibitem[Feynman(1982)]%
        {Feynman1982}
\bibfield{author}{\bibinfo{person}{Richard~P. Feynman}.}
  \bibinfo{year}{1982}\natexlab{}.
\newblock \showarticletitle{Simulating physics with computers}.
\newblock \bibinfo{journal}{\emph{International Journal of Theoretical
  Physics}} \bibinfo{volume}{21}, \bibinfo{number}{6} (\bibinfo{year}{1982}),
  \bibinfo{pages}{467--488}.
\newblock
\urldef\tempurl%
\url{https://doi.org/10.1007/BF02650179}
\showDOI{\tempurl}


\bibitem[{Google Quantum AI}(2023)]%
        {Qsim}
\bibfield{author}{\bibinfo{person}{{Google Quantum AI}}.}
  \bibinfo{year}{2023}\natexlab{}.
\newblock \bibinfo{title}{qsim: {Google}'s Quantum Circuit Simulator}.
\newblock
\newblock
\urldef\tempurl%
\url{https://github.com/quantumlib/qsim}
\showURL{%
\tempurl}


\bibitem[{Google Quantum AI}(2024)]%
        {Cirq2024}
\bibfield{author}{\bibinfo{person}{{Google Quantum AI}}.}
  \bibinfo{year}{2024}\natexlab{}.
\newblock \bibinfo{title}{Cirq: A Python Framework for Quantum Circuits}.
\newblock
\newblock
\urldef\tempurl%
\url{https://quantumai.google/cirq}
\showURL{%
\tempurl}


\bibitem[Grover(1996)]%
        {grover1996fast}
\bibfield{author}{\bibinfo{person}{Lov~K Grover}.}
  \bibinfo{year}{1996}\natexlab{}.
\newblock \showarticletitle{A fast quantum mechanical algorithm for database
  search}. In \bibinfo{booktitle}{\emph{Proceedings of the 28th Annual {ACM}
  Symposium on Theory of Computing ({STOC})}}. \bibinfo{pages}{212--219}.
\newblock
\urldef\tempurl%
\url{https://doi.org/10.1145/237814.237866}
\showDOI{\tempurl}


\bibitem[H{\"a}ner and Steiger(2017)]%
        {HanerSteiger2017}
\bibfield{author}{\bibinfo{person}{Thomas H{\"a}ner} {and}
  \bibinfo{person}{Damian~S. Steiger}.} \bibinfo{year}{2017}\natexlab{}.
\newblock \showarticletitle{0.5 petabyte simulation of a 45-qubit quantum
  circuit}. In \bibinfo{booktitle}{\emph{Proceedings of the International
  Conference for High Performance Computing, Networking, Storage and Analysis
  (SC '17)}}.
\newblock
\urldef\tempurl%
\url{https://doi.org/10.1145/3126908.3126947}
\showDOI{\tempurl}


\bibitem[{Intel Corporation}(2023)]%
        {IntelQS}
\bibfield{author}{\bibinfo{person}{{Intel Corporation}}.}
  \bibinfo{year}{2023}\natexlab{}.
\newblock \bibinfo{title}{Intel Quantum Simulator}.
\newblock
\newblock
\urldef\tempurl%
\url{https://github.com/intel/intel-qs}
\showURL{%
\tempurl}


\bibitem[Jones et~al\mbox{.}(2019)]%
        {Jones2019quest}
\bibfield{author}{\bibinfo{person}{Tyson Jones}, \bibinfo{person}{Anna Brown},
  \bibinfo{person}{Ian Bush}, {and} \bibinfo{person}{Simon~C. Benjamin}.}
  \bibinfo{year}{2019}\natexlab{}.
\newblock \showarticletitle{{QuEST} and High Performance Simulation of Quantum
  Computers}.
\newblock \bibinfo{journal}{\emph{Scientific Reports}}  \bibinfo{volume}{9}
  (\bibinfo{year}{2019}), \bibinfo{pages}{10736}.
\newblock
\urldef\tempurl%
\url{https://doi.org/10.1038/s41598-019-47174-9}
\showDOI{\tempurl}


\bibitem[Li et~al\mbox{.}(2023)]%
        {QASMBench2020}
\bibfield{author}{\bibinfo{person}{Ang Li}, \bibinfo{person}{Samuel Stein},
  \bibinfo{person}{Sriram Krishnamoorthy}, {and} \bibinfo{person}{James Ang}.}
  \bibinfo{year}{2023}\natexlab{}.
\newblock \showarticletitle{{QASMBench}: A Low-Level Quantum Benchmark Suite
  for {NISQ} Evaluation and Simulation}.
\newblock \bibinfo{journal}{\emph{{ACM} Transactions on Quantum Computing}}
  \bibinfo{volume}{4}, \bibinfo{number}{2} (\bibinfo{year}{2023}),
  \bibinfo{pages}{10:1--10:26}.
\newblock
\urldef\tempurl%
\url{https://doi.org/10.1145/3550488}
\showDOI{\tempurl}


\bibitem[Lloyd(1996)]%
        {Lloyd1996}
\bibfield{author}{\bibinfo{person}{Seth Lloyd}.}
  \bibinfo{year}{1996}\natexlab{}.
\newblock \showarticletitle{Universal quantum simulators}.
\newblock \bibinfo{journal}{\emph{Science}} \bibinfo{volume}{273},
  \bibinfo{number}{5278} (\bibinfo{year}{1996}), \bibinfo{pages}{1073--1078}.
\newblock
\urldef\tempurl%
\url{https://doi.org/10.1126/science.273.5278.1073}
\showDOI{\tempurl}


\bibitem[Luo et~al\mbox{.}(2020)]%
        {YaoJL2024}
\bibfield{author}{\bibinfo{person}{Xiu-Zhe Luo}, \bibinfo{person}{Jin-Guo Liu},
  \bibinfo{person}{Pan Zhang}, {and} \bibinfo{person}{Lei Wang}.}
  \bibinfo{year}{2020}\natexlab{}.
\newblock \showarticletitle{Yao.jl: Extensible, Efficient Framework for Quantum
  Algorithm Design}.
\newblock \bibinfo{journal}{\emph{Quantum}}  \bibinfo{volume}{4}
  (\bibinfo{year}{2020}), \bibinfo{pages}{341}.
\newblock
\urldef\tempurl%
\url{https://doi.org/10.22331/q-2020-10-11-341}
\showDOI{\tempurl}


\bibitem[Nation et~al\mbox{.}(2025)]%
        {Benchpress2024}
\bibfield{author}{\bibinfo{person}{Paul~D. Nation}, \bibinfo{person}{Abdullah
  {Ash Saki}}, \bibinfo{person}{Sebastian Brandhofer}, \bibinfo{person}{Luciano
  Bello}, \bibinfo{person}{Shelly Garion}, \bibinfo{person}{Matthew Treinish},
  {and} \bibinfo{person}{Ali Javadi-Abhari}.} \bibinfo{year}{2025}\natexlab{}.
\newblock \showarticletitle{Benchmarking the performance of quantum computing
  software}.
\newblock \bibinfo{journal}{\emph{Nature Computational Science}}
  (\bibinfo{year}{2025}).
\newblock
\showeprint[arxiv]{2409.08844}
\urldef\tempurl%
\url{https://arxiv.org/abs/2409.08844}
\showURL{%
\tempurl}


\bibitem[Nielsen and Chuang(2010)]%
        {NielsenChuang2010}
\bibfield{author}{\bibinfo{person}{Michael~A Nielsen} {and}
  \bibinfo{person}{Isaac~L Chuang}.} \bibinfo{year}{2010}\natexlab{}.
\newblock \bibinfo{booktitle}{\emph{Quantum Computation and Quantum
  Information: 10th Anniversary Edition}}.
\newblock \bibinfo{publisher}{Cambridge University Press}.
\newblock


\bibitem[Peruzzo et~al\mbox{.}(2014)]%
        {peruzzo2014variational}
\bibfield{author}{\bibinfo{person}{Alberto Peruzzo}, \bibinfo{person}{Jarrod
  McClean}, \bibinfo{person}{Peter Shadbolt}, \bibinfo{person}{Man-Hong Yung},
  \bibinfo{person}{Xiao-Qi Zhou}, \bibinfo{person}{Peter~J Love},
  \bibinfo{person}{Al{\'a}n Aspuru-Guzik}, {and} \bibinfo{person}{Jeremy~L
  O'Brien}.} \bibinfo{year}{2014}\natexlab{}.
\newblock \showarticletitle{A variational eigenvalue solver on a photonic
  quantum processor}.
\newblock \bibinfo{journal}{\emph{Nature Communications}} \bibinfo{volume}{5},
  \bibinfo{number}{1} (\bibinfo{year}{2014}), \bibinfo{pages}{4213}.
\newblock
\urldef\tempurl%
\url{https://doi.org/10.1038/ncomms5213}
\showDOI{\tempurl}


\bibitem[Preskill(2018)]%
        {Preskill2018}
\bibfield{author}{\bibinfo{person}{John Preskill}.}
  \bibinfo{year}{2018}\natexlab{}.
\newblock \showarticletitle{Quantum Computing in the {NISQ} era and beyond}.
\newblock \bibinfo{journal}{\emph{Quantum}}  \bibinfo{volume}{2}
  (\bibinfo{year}{2018}), \bibinfo{pages}{79}.
\newblock
\urldef\tempurl%
\url{https://doi.org/10.22331/q-2018-08-06-79}
\showDOI{\tempurl}


\bibitem[{QTechTheory, University of Oxford}(2023)]%
        {QuEST}
\bibfield{author}{\bibinfo{person}{{QTechTheory, University of Oxford}}.}
  \bibinfo{year}{2023}\natexlab{}.
\newblock \bibinfo{title}{{QuEST}: Quantum Exact Simulation Toolkit}.
\newblock
\newblock
\urldef\tempurl%
\url{https://github.com/QuEST-Kit/QuEST}
\showURL{%
\tempurl}


\bibitem[Quetschlich et~al\mbox{.}(2023)]%
        {MQTBench2022}
\bibfield{author}{\bibinfo{person}{Nils Quetschlich}, \bibinfo{person}{Lukas
  Burgholzer}, {and} \bibinfo{person}{Robert Wille}.}
  \bibinfo{year}{2023}\natexlab{}.
\newblock \showarticletitle{{MQT} Bench: Benchmarking Software and Design
  Automation Tools for Quantum Computing}.
\newblock \bibinfo{journal}{\emph{Quantum}}  \bibinfo{volume}{7}
  (\bibinfo{year}{2023}), \bibinfo{pages}{1062}.
\newblock
\urldef\tempurl%
\url{https://doi.org/10.22331/q-2023-07-20-1062}
\showDOI{\tempurl}
\showeprint[arxiv]{2204.13719}


\bibitem[{QuTiP collaboration}(2025)]%
        {QuantumToolboxJulia2024}
\bibfield{author}{\bibinfo{person}{{QuTiP collaboration}}.}
  \bibinfo{year}{2025}\natexlab{}.
\newblock \bibinfo{title}{{QuantumToolbox.jl}: An efficient {Julia} framework
  for simulating open quantum systems}.
\newblock
\newblock
\showeprint[arxiv]{2504.21440}
\urldef\tempurl%
\url{https://github.com/qutip/QuantumToolbox.jl}
\showURL{%
\tempurl}


\bibitem[Schieffer et~al\mbox{.}(2024)]%
        {UnifiedMemory2024}
\bibfield{author}{\bibinfo{person}{Gabin Schieffer}, \bibinfo{person}{Jacob
  Wahlgren}, \bibinfo{person}{Jie Ren}, \bibinfo{person}{Jennifer Faj}, {and}
  \bibinfo{person}{Ivy Peng}.} \bibinfo{year}{2024}\natexlab{}.
\newblock \showarticletitle{Harnessing Integrated {CPU-GPU} System Memory for
  {HPC}: a first look into {Grace Hopper}}.
\newblock \bibinfo{journal}{\emph{Proceedings of the 53rd International
  Conference on Parallel Processing (ICPP)}} (\bibinfo{year}{2024}).
\newblock
\urldef\tempurl%
\url{https://doi.org/10.1145/3673038.3673110}
\showDOI{\tempurl}
\showeprint[arxiv]{2407.07850}


\bibitem[Schleich et~al\mbox{.}(2024)]%
        {QuantumSimulation2025}
\bibfield{author}{\bibinfo{person}{Philipp Schleich}, \bibinfo{person}{Jakob~S.
  Kottmann}, {and} \bibinfo{person}{Al{\'a}n Aspuru-Guzik}.}
  \bibinfo{year}{2024}\natexlab{}.
\newblock \showarticletitle{Simulation of Quantum Computers: Review and
  Acceleration Opportunities}.
\newblock \bibinfo{journal}{\emph{arXiv preprint arXiv:2410.12660}}
  (\bibinfo{year}{2024}).
\newblock


\bibitem[Smith and Gray(2018)]%
        {SmithGray2018}
\bibfield{author}{\bibinfo{person}{Daniel G.~A. Smith} {and}
  \bibinfo{person}{Johnnie Gray}.} \bibinfo{year}{2018}\natexlab{}.
\newblock \showarticletitle{opt\_einsum---{A} {Python} package for optimizing
  contraction order for {Einstein} summation expressions}.
\newblock \bibinfo{journal}{\emph{Journal of Open Source Software}}
  \bibinfo{volume}{3}, \bibinfo{number}{26} (\bibinfo{year}{2018}),
  \bibinfo{pages}{753}.
\newblock
\urldef\tempurl%
\url{https://doi.org/10.21105/joss.00753}
\showDOI{\tempurl}


\bibitem[Suzuki(1976)]%
        {Suzuki1976}
\bibfield{author}{\bibinfo{person}{Masuo Suzuki}.}
  \bibinfo{year}{1976}\natexlab{}.
\newblock \showarticletitle{Generalized {T}rotter's formula and systematic
  approximants of exponential operators and inner derivations with applications
  to many-body problems}.
\newblock \bibinfo{journal}{\emph{Communications in Mathematical Physics}}
  \bibinfo{volume}{51}, \bibinfo{number}{2} (\bibinfo{year}{1976}),
  \bibinfo{pages}{183--190}.
\newblock
\urldef\tempurl%
\url{https://doi.org/10.1007/BF01609348}
\showDOI{\tempurl}


\bibitem[Suzuki et~al\mbox{.}(2021)]%
        {Qulacs}
\bibfield{author}{\bibinfo{person}{Yasunari Suzuki}, \bibinfo{person}{Yoshiaki
  Kawase}, \bibinfo{person}{Yuya Masumura}, \bibinfo{person}{Yuria Hiraga},
  \bibinfo{person}{Masahiro Nakadai}, \bibinfo{person}{Jiabao Chen},
  \bibinfo{person}{Ken~M. Nakanishi}, \bibinfo{person}{Kosuke Mitarai},
  \bibinfo{person}{Ryosuke Imai}, \bibinfo{person}{Shiro Tamiya},
  \bibinfo{person}{Takahiro Yamamoto}, \bibinfo{person}{Tennin Yan},
  \bibinfo{person}{Toru Kawakubo}, \bibinfo{person}{Yuya~O. Nakagawa},
  \bibinfo{person}{Yohei Ibe}, \bibinfo{person}{Youyuan Zhang},
  \bibinfo{person}{Hirotsugu Yamashita}, \bibinfo{person}{Hikaru Yoshimura},
  \bibinfo{person}{Akihiro Hayashi}, {and} \bibinfo{person}{Keisuke Fujii}.}
  \bibinfo{year}{2021}\natexlab{}.
\newblock \showarticletitle{Qulacs: a fast and versatile quantum circuit
  simulator for research purpose}.
\newblock \bibinfo{journal}{\emph{Quantum}}  \bibinfo{volume}{5}
  (\bibinfo{year}{2021}), \bibinfo{pages}{559}.
\newblock
\urldef\tempurl%
\url{https://doi.org/10.22331/q-2021-10-06-559}
\showDOI{\tempurl}


\bibitem[Tomesh et~al\mbox{.}(2022)]%
        {SupermarQ2022}
\bibfield{author}{\bibinfo{person}{Teague Tomesh}, \bibinfo{person}{Pranav
  Gokhale}, \bibinfo{person}{Victory Omole}, \bibinfo{person}{Gokul~Subramanian
  Ravi}, \bibinfo{person}{Kaitlin~N. Smith}, \bibinfo{person}{Joshua Viszlai},
  \bibinfo{person}{Xin-Chuan Wu}, \bibinfo{person}{Nikos Hardavellas},
  \bibinfo{person}{Margaret~R. Martonosi}, {and} \bibinfo{person}{Frederic~T.
  Chong}.} \bibinfo{year}{2022}\natexlab{}.
\newblock \showarticletitle{{SupermarQ}: A Scalable Quantum Benchmark Suite}.
  In \bibinfo{booktitle}{\emph{2022 {IEEE} International Symposium on
  High-Performance Computer Architecture ({HPCA})}}. \bibinfo{pages}{587--603}.
\newblock
\urldef\tempurl%
\url{https://doi.org/10.1109/HPCA53966.2022.00050}
\showDOI{\tempurl}
\showeprint[arxiv]{2202.11045}


\end{thebibliography}

\appendix

\section{Pure MLX versus a Hand-Written Kernel, Line by Line}\label{app:hlayer}

This appendix walks through ONE case end to end, so that the difference between the two tiers is concrete rather than statistical: applying a Hadamard gate to every qubit of a 25-qubit state. This layer appears at the start of Grover search, in QAOA state preparation, in Deutsch--Jozsa, and (twice per Trotter step) inside the basis conjugations of the XX and YY layers of Section~\ref{sec:shaders}. The state holds $2^{25}$ complex64 amplitudes, 268\,MB; every full read-plus-write of it moves 0.537\,GB.

\subsection{The pure-MLX path}

The pure tier routes each H gate through the structured single-qubit update:

\begin{lstlisting}[language=Python,caption={Pure MLX: one single-qubit gate (src/mlxq/sim.py)},label={lst:mlx-h}]
def apply_single(self, gate, q):
    ax = canonical_axis_index(q, self.n)
    # View the state as (A, 2, B): axis `ax`
    # isolates qubit q's bit.
    a_dim = 1 << ax
    b_dim = 1 << (self.n - ax - 1)
    tensor = mx.reshape(self.state,
                        (a_dim, 2, b_dim))
    s0 = tensor[:, 0, :]   # amplitudes with bit=0
    s1 = tensor[:, 1, :]   # amplitudes with bit=1
    n0 = gate[0,0]*s0 + gate[0,1]*s1
    n1 = gate[1,0]*s0 + gate[1,1]*s1
    self.state = mx.reshape(
        mx.stack([n0, n1], axis=1), (1 << self.n,))
\end{lstlisting}

Step by step: line 3 computes which axis of the $(A, 2, B)$ view corresponds to qubit $q$ (MSB-first, so qubit $q$ is bit $n{-}1{-}q$). Lines 7--10 slice the state into the half where that bit is 0 and the half where it is 1; no data moves yet, because MLX slices are lazy views. Lines 11--12 form the new halves as a general $2\times2$ complex linear combination: four scalar-vector complex multiplies and two adds, correct for ANY single-qubit gate. Line 13 reassembles the state. When MLX evaluates this graph it reads the full state once and writes it once: $0.537$\,GB of traffic, plus graph construction and scheduling overhead for the five array operations involved.

An all-qubit H layer calls this 25 times, once per qubit. Nothing is wrong with any single call; the cost is structural: $25$ full-state passes ($\approx 13.4$\,GB of traffic), $25$ lazy-graph constructions, and a full $2\times2$ complex combination per gate even though H's entries are all $\pm 1/\sqrt{2}$. Measured on an idle M1 Max, the structured per-gate layer takes $187$\,ms --- the same cost as the structurally identical RX layer (Table~\ref{tab:shader}), as expected since both apply 25 sequential single-qubit updates.

\subsection{The Metal path}

The shader tier fuses the layer. Because single-qubit gates on distinct wires commute, H on 25 qubits may be regrouped freely; the kernel applies $H^{\otimes 4}$ to four adjacent qubits per pass. For four qubits, the tensor-product matrix has entries $\pm 1/4$ with the sign given by a parity: $(H^{\otimes 4})_{rc} = \tfrac{1}{4}(-1)^{\mathrm{popcount}(r \wedge c)}$. That closed form is the entire kernel:

\begin{lstlisting}[language=C++,caption={Metal: $H^{\otimes4}$ on four adjacent qubits, one thread per 16 amplitudes (src/mlxq/shaders/single\_qubit.py)},label={lst:metal-h}]
uint q = thread_position_in_grid.x;
if (q >= n_hexads) return;          // bounds guard
uint bit_low = 1u << (shift_hi - 3u);
uint low  = q & (bit_low - 1u);     // bits below block
uint high = q >> (shift_hi - 3u);   // bits above block
uint base = (high << (shift_hi + 1u)) | low;

complex64_t a[16];  uint idx[16];
for (uint c = 0; c < 16u; ++c) {    // load tile
    idx[c] = base | (c << (shift_hi - 3u));
    a[c] = state[idx[c]];
}
const float s = 0.25f;
for (uint r = 0; r < 16u; ++r) {    // H (x) H (x) H (x) H
    float accr = 0.0f, acci = 0.0f;
    for (uint c = 0; c < 16u; ++c) {
        if (metal::popcount(r & c) & 1u)
             { accr -= a[c].real; acci -= a[c].imag; }
        else { accr += a[c].real; acci += a[c].imag; }
    }
    out[idx[r]] = complex64_t(accr*s, acci*s);
}
\end{lstlisting}

Step by step. Lines 1--2: one thread owns one 16-amplitude tile; the guard makes over-dispatch safe. Lines 3--6: index arithmetic splits the basis index into the bits below and above the 4-qubit block, exactly as the $(A, 2, B)$ view did for one qubit, except the ``2'' is now ``16''. Lines 8--12: the tile is loaded into registers; these are the only global reads the pass performs. Lines 13--22: the $16\times16$ Walsh matrix is never materialized; each output is an accumulation of the 16 inputs with a sign from one popcount, and the single multiply by $0.25$ applies all four $1/\sqrt{2}$ factors at once. Every amplitude is read once and written once PER PASS, regardless of the four gates the pass absorbed.

The host launches $\lceil 25/4 \rceil = 7$ passes (six radix-4 blocks and one single-qubit tail). Total traffic: $7 \times 0.537 \approx 3.8$\,GB.

\subsection{Where the $\sim$8$\times$ comes from}

Measured, the fused all-H layer drops from $187$ to $23.7$\,ms (Table~\ref{tab:shader}), a $\sim$8$\times$ reduction.\footnote{This is a clean idle-machine re-measurement (3 warmups, 10 repeats, machine-quietness probe; \path{evidence_artifacts/h_layer_clean_20260706.json}). An earlier draft reported a $698$\,ms per-gate baseline and a $35\times$ (originally $55\times$) speedup; a controlled re-measurement did not reproduce that baseline. The per-gate cost of an all-H layer is the same as the structurally identical RX layer ($\sim$190\,ms), so the honest fused speedup is $\sim$8$\times$, not an order of magnitude. The $12.7$\,ms figure some intermediate notes carried was a single best-case run and is not campaign-grade.} Pass counting explains a factor of $25/7 \approx 3.6$: the pure path makes 25 sequential passes, the Walsh kernel makes 7. The remaining ${\approx}2.2\times$ is what fusion removes per pass --- 25 lazy-graph constructions and synchronization points collapse into 7 kernel launches, and the $2\times2$ complex combination becomes sign-flipped adds. The end state is a bandwidth argument: $3.8$\,GB in $23.7$\,ms sustains ${\approx}160$\,GB/s, about $40\%$ of the M1 Max's 400\,GB/s peak, so the kernel is already bandwidth-bound while the pure path runs close to the framework's dispatch overhead. The same accounting predicts where hand-tuning does NOT pay: the ZZ Trotter layer is one broadcast multiply on the pure path already (Table~\ref{tab:shader}), and its kernel gains only $1.5\times$; and it explains why H, lacking a diagonal shortcut, sees a single-digit rather than order-of-magnitude gain.

\section{Complete Per-Workload Speedups}\label{app:sweep}

Table~\ref{tab:sweep-full} reports the shader tier against the pure-MLX tier for EVERY gate-based workload in the benchmark suite at 25 qubits, from the paired sweep (\path{tools/shader_suite_sweep.py}): within each repeat the pure and Metal arms alternate, so session drift hits both equally, and the ratio column is the mean of per-repeat paired ratios with its observed range. Excluded are \path{steady_state} (density-matrix Kraus path, no gate list) and \path{qft_fft_primitive} (an algorithmic primitive, not gate execution).

\begin{table}[t]
\caption{Full-suite paired sweep at 25 qubits: pure MLX vs.\ Metal shaders (five paired repeats after per-arm warmup; ratio column is the mean of per-repeat paired ratios with the observed range; sorted by speedup). Artifact: \texttt{evidence\_artifacts/shader\_sweep\_20260704/}.}
\label{tab:sweep-full}
\centering
\scriptsize
\setlength{\tabcolsep}{3.5pt}
\begin{tabular}{@{}lrrrc@{}}
\toprule
Workload & Pure MLX & Metal & Speedup & Range \\
\midrule
TFIM Trotter (2nd order) & 15.8\,s & 632.4\,ms & $25.0\times$ & $[22.7, 26.9]$ \\
Long-range Ising & 14.8\,s & 614.9\,ms & $24.3\times$ & $[20.3, 26.4]$ \\
Variational ansatz & 2.14\,s & 158.7\,ms & $14.3\times$ & $[10.3, 21.7]$ \\
Heisenberg chain & 23.1\,s & 1.71\,s & $13.8\times$ & $[11.2, 16.4]$ \\
Heisenberg XXZ & 23.1\,s & 1.78\,s & $13.2\times$ & $[11.5, 15.9]$ \\
cuQuantum/blueqat proxy & 6.16\,s & 469.5\,ms & $13.2\times$ & $[11.3, 14.7]$ \\
TFIM random field & 7.83\,s & 600.7\,ms & $13.1\times$ & $[12.3, 13.7]$ \\
TFIM Trotter (1st order) & 6.91\,s & 567.1\,ms & $12.3\times$ & $[10.6, 13.7]$ \\
QCBM & 1.94\,s & 198.0\,ms & $12.0\times$ & $[5.2, 22.3]$ \\
Quantum walk (v-chain) & 4.54\,s & 395.3\,ms & $11.6\times$ & $[10.3, 15.2]$ \\
Grover search & 1.05\,s & 131.4\,ms & $11.5\times$ & $[5.1, 20.0]$ \\
Quantum walk & 4.52\,s & 401.7\,ms & $11.4\times$ & $[8.7, 12.7]$ \\
EfficientSU2 (random) & 2.62\,s & 249.3\,ms & $10.6\times$ & $[7.7, 13.6]$ \\
GHZ preparation & 210.1\,ms & 23.4\,ms & $9.1\times$ & $[6.3, 11.9]$ \\
Heisenberg random field & 24.7\,s & 2.78\,s & $8.9\times$ & $[8.2, 10.2]$ \\
Graph state & 281.0\,ms & 31.5\,ms & $8.9\times$ & $[8.4, 9.4]$ \\
Phase estimation (inexact) & 484.8\,ms & 55.0\,ms & $8.8\times$ & $[8.3, 10.0]$ \\
QFT & 702.1\,ms & 108.8\,ms & $8.4\times$ & $[3.1, 11.7]$ \\
QAOA (ring) & 1.57\,s & 215.5\,ms & $8.4\times$ & $[4.9, 15.9]$ \\
Deutsch--Jozsa & 418.0\,ms & 50.0\,ms & $8.4\times$ & $[7.9, 9.0]$ \\
Phase estimation & 924.6\,ms & 158.9\,ms & $8.0\times$ & $[2.3, 11.0]$ \\
QNN layer & 740.8\,ms & 119.1\,ms & $7.4\times$ & $[3.1, 8.8]$ \\
QFT (entangled input) & 890.0\,ms & 150.0\,ms & $7.3\times$ & $[3.6, 12.9]$ \\
RealAmplitudes & 970.4\,ms & 237.3\,ms & $5.1\times$ & $[2.5, 10.0]$ \\
Random circuit & 1.41\,s & 298.1\,ms & $4.7\times$ & $[4.0, 5.8]$ \\
VQE ansatz + energy & 2.00\,s & 1.66\,s & $1.3\times$ & $[0.9, 2.0]$ \\
Heisenberg 2-leg ladder & 19.5\,s & 19.2\,s & $1.0\times$ & $[1.0, 1.1]$ \\
Amplitude estimation & 364.7\,ms & 365.1\,ms & $1.0\times$ & $[1.0, 1.0]$ \\
W state & 323.1\,ms & 326.5\,ms & $1.0\times$ & $[1.0, 1.0]$ \\
\bottomrule
\end{tabular}
\end{table}

Twenty-five of the 29 workloads accelerate with the observed paired-ratio range (five repeats) entirely above parity, by $4.7$--$25\times$ where fusable layer structure exists; unlike the four-way campaign (ten repeats, t-based CIs), this sweep reports ranges rather than confidence intervals. Four do not clear that bar, each for a documented structural reason. VQE has a $1.3\times$ point estimate but an interval spanning parity, because its residual cost is expectation-value evaluation rather than gate execution. The two-leg Heisenberg ladder ($1.02\times$, interval spanning parity) interleaves XX, YY, and ZZ terms per bond, so no same-family run exists and reordering across families would change the Trotterized operator. Amplitude estimation ($1.00\times$) places single CZ gates between short single-qubit blocks, leaving no run of length two, and W-state preparation ($0.99\times$) strictly alternates RY and CNOT gates. These rows quantify the boundary of the fusion approach: the shader tier accelerates layer STRUCTURE, and circuits without it correctly fall through to the pure path at a fractional-percent overhead.

\subsection*{B.1 \quad The OpenQASM import path}

The tables above use circuits built by our own benchmark generators, so a separate question remains: does the shader tier engage on EXTERNALLY authored circuits arriving through OpenQASM~2.0 import, where gate vocabulary (\texttt{cx}, \texttt{cu1}, \texttt{p}, \texttt{ccx}), barriers, and decomposition choices are outside our control? We executed the full local QASM corpus (42 files, QASMBench-derived and hand-written; \path{tools/qasm_shader_sweep.py}, three paired repeats, two files skipped under a documented cost cap and logged in the manifest) through both tiers. All 40 executed circuits agree between the tiers to $\le 5\times10^{-7}$. Representative timings (mean ms):

\begin{center}
\footnotesize
\begin{tabular}{@{}lrrrr@{}}
\toprule
Circuit & Qubits & Pure MLX & Metal & Ratio \\
\midrule
ghz\_state\_n23 & 23 & 36.9 & 4.1 & $9.1\times$ \\
cat\_state\_n22 & 22 & 18.9 & 2.4 & $8.0\times$ \\
knn\_n25 & 25 & 344.2 & 125.5 & $2.7\times$ \\
swap\_test\_n25 & 25 & 312.0 & 123.8 & $2.5\times$ \\
ising\_n26 & 26 & 2382.6 & 1074.1 & $2.2\times$ \\
wstate\_n27 & 27 & 3047.1 & 2222.5 & $1.4\times$ \\
qft\_n18 & 18 & 33.5 & 34.7 & $0.96\times$ \\
factor247\_n15 & 15 & 23102 & 22897 & $1.01\times$ \\
\bottomrule
\end{tabular}
\end{center}

The import path normalizes gate names into the detector vocabulary (\texttt{cx}$\to$CNOT, \texttt{cu1}/\texttt{cp}$\to$CPHASE) and skips barriers, so structured circuits fuse exactly as the native ones do: the GHZ and cat-state CNOT chains collapse to one gather each. Two instructive non-wins: QASMBench's \texttt{qft\_n18} arrives PRE-DECOMPOSED into \texttt{cx}+\texttt{rz} sequences rather than controlled-phase gates, so no CPHASE ladder exists in the stream for the stage detector to recognize, and \texttt{factor247\_n15} is 660{,}566 gates of dense irregular arithmetic with no layer structure. Both fall through to the pure path at parity ($0.96\times$--$1.01\times$) rather than mis-fusing; recognizing decomposed two-qubit blocks (Aer-style generic matrix fusion) is the natural next extension and is noted as future work. Circuits below ${\sim}15$ qubits sit at kernel-launch overhead and show ratios near $1\times$ regardless of structure. Artifact: \path{evidence_artifacts/qasm_shader_sweep_20260704/}.

\end{document}